\documentclass[%
 aip,
 jmp,%
 amsmath,amssymb,
 reprint,%
]{revtex4-1}

\usepackage{graphicx}% Include figure files
\usepackage{dcolumn}% Align table columns on decimal point
\usepackage{bm}% bold math
\usepackage{epstopdf}

\begin{document}

\preprint{AIP/123-QED}

\title[Phys.\ Plasmas {\bf{20}}, 122310 (2013); doi: 10.1063/1.4851976]{Transport enhancement and suppression in turbulent magnetic reconnection: A self-consistent turbulence model}% Force line breaks with \\
%\thanks{Footnote to title of article.}

\author{N. Yokoi}
 \altaffiliation[]{Guest researcher at the National Astronomical Observatory of Japan (NAOJ) and the Nordic Institute for Theoretical Physics (NORDITA).}%Lines break automatically or can be forced with \\
\affiliation{ 
Institute of Industrial Science, University of Tokyo%\\This line break forced with \textbackslash\textbackslash
}%
\email{nobyokoi@iis.u-tokyo.ac.jp.}

\author{K. Higashimori}%
\affiliation{%
Department of Earth and Planetary Science, University of Tokyo%\\This line break forced% with \\
}%

\author{M. Hoshino}
\affiliation{%
Department of Earth and Planetary Science, University of Tokyo%\\This line break forced% with \\
}%

\date{received 7 February 2013, accepted 29 November 2013}

\begin{abstract}
Through the enhancement of transport, turbulence is expected to contribute to the fast reconnection. However the effects of turbulence are not so straightforward. In addition to the enhancement of transport, turbulence under some environment shows effects that suppress the transport. In the presence of turbulent cross helicity, such a dynamic balance between the transport enhancement and suppression occurs. As this result of dynamic balance, the region of effective enhanced magnetic diffusivity is confined to a narrow region, leading to the fast reconnection. In order to confirm this idea, a self-consistent turbulence model for the magnetic reconnection is proposed. With the aid of numerical simulations where turbulence effects are incorporated in a consistent manner through the turbulence model, the dynamic balance in the turbulence magnetic reconnection is confirmed.
\end{abstract}

%Use showkeys class option if keyword
\maketitle

%-----------------------------------------------------------------------------
%	I Introduction
%-----------------------------------------------------------------------------
\section{\label{sec:level1}Introduction}
The concept of magnetic reconnection is important to understand a variety of plasma phenomena observed in astrophysics, space physics, fusion plasma physics from a unified viewpoint. Aurorae are believed to be related to the magnetic reconnection in the Earth's magnetosphere. Extremely large amount of energy release in solar flare is considered to be due to the reconnection. Several jets ubiquitously observed in astrophysical objects may be caused by the magnetic reconnection. Reconnection of the magnetic field has also been proposed as a source of the viscous dissipation in the accretion disk. 

	A steady state of magnetic reconnection was investigated by Sweet and Parker with the aid of a simple magnetohydrodynamic (MHD) model.\cite{swe1958,par1957}  However, their model was found to be not sufficient when it is applied to solar flares. Most obvious problems in the theory of magnetic reconnection have been how to realize a fast reconnection and how to bridge the gap of scales. Since then several researchers including Petschek have tackled this fast magnetic reconnection mechanism.\cite{pet1964} The reconnection rate or inflow Alfv\'{e}n Mach Number $M_{\rm{in}}$ in the Sweet--Parker mechanism is written as $M_{\rm{i}} \equiv U_{\rm{in}} / V_{\rm{Ain}} = \delta / L_0 = Lu^{-1/2}$, where $U_{\rm{in}}$ is the inflow speed, $V_{\rm{Ain}}$ the inflow Alfv\'{e}n speed, $\delta$ the thickness of the diffusion region, $L_0$ the length scale of phenomena, $Lu$ the Lundquist number (the magnetic Reynolds number based on the inflow Alfv\'{e}n speed). In the astrophysical and space plasma situations, the Lundquist number is usually huge ($Lu \gg 10^{6}$), so the reconnection rate becomes very small with the Sweet--Parker model. In other words, magnetic reconnection model applied to the astrophysical and space plasma situations is too slow to explain the real phenomena such as the solar flare. Another problem is the gap of scales. In the solar flare case, the typical scale of phenomena is $L_0 \sim 10^4\ {\rm{km}}$. On the other hand, the thickness of diffusion region or current sheet is $\delta = \rho_{\rm{i}} \sim 10\ {\rm{m}}$ ($\rho_{\rm{i}}$: ion-Lamor radius). This means that the dynamics of very huge scale motions are determined by that of very small-scale behaviors of the magnetic field. In order to get a realistic view of the solar flare phenomena, we need highly localized effective resistivity, and also something that match this scale gap. 

	Several mechanisms that potentially elucidate the fast and localized reconnection through the enhanced resistivity have been investigated, which include the electron inertia,\cite{vas1975} Hall effect,\cite{ter1983,bir2001} electron pressure,\cite{vas1975,hes1998} ambipolar diffusion,\cite{sin2011} kinetic effect,\cite{buc1999,hig2013b} etc. Turbulence is also considered as one of such effects. A prominent feature of turbulence is multiple-scale interaction of motions. Due to the nonlinear term in Navier--Stokes equation, the dynamics of  motion represented by wave number mode $\bf{k}$ is governed by its interaction with all the other scales of motions. Another prominent feature is that turbulence enhances transport of the system very much. The representative effects are the eddy viscosity, eddy diffusivity, turbulent magnetic diffusivity, etc. In the case of magnetic diffusivity, the molecular diffusivity $\eta$ is enhanced by turbulence as $\eta \to \eta + \beta$, where $\beta$ is the turbulent magnetic diffusivity. The ratio of the turbulent to molecular magnetic diffusivities is written as $\beta / \eta \sim Rm^{({\rm{T}})}$, where $Rm^{({\rm{T}})}$ is the turbulent magnetic Reynolds number. In astrophysical and space physics phenomena, the magnetic Reynolds number is huge. So, the turbulent counterpart is also expected to be very large. This means that the effective magnetic diffusivity due to turbulence is much larger than the molecular one ($\beta \gg \eta$). 
We should note that, in the context of mean-field magnetic reconnection, the large effective magnetic diffusivity does not necessarily lead to the fast and enhanced reconnection. This has been recently shown numerically with magnetohydrodynamic (MHD) turbulence model simulations.\cite{hig2013a} We will return to this point later in \S~V.C. If the magnetic field structure is diffused everywhere by the enhanced magnetic diffusivity, we just have diffusion or annihilation of the oppositely directed magnetic field in a broad region. There it is difficult to generate open fast reconnection jets from the mean magnetic field. In order to get a fast reconnection, the region of high diffusivity should be confined to a very small region.

	Since turbulence is expected to play an important role in enhancing the magnetic reconnection rate and also to bridge the scale gap between the the dissipation scale and the scale of phenomena,\cite{hos1994,taj1997} over the past few decades, a considerable number of studies have been made on turbulent reconnection. In the context of fast reconnection, we should note that the magnetic reconnection rate is basically determined by the ratio of the thickness and width of the reconnection region. In order to get a fast reconnection, we have two ways: (i) increasing the effective thickness due to the fluctuating motion of the magnetic field; (ii) decreasing the effective width by generating multiple magnetic islands (fractal structure, plasmoids, etc.).\cite{kar2013}
	
	One of the earliest investigations on the turbulence effects on magnetic reconnection was made by Matthaeus and Lamkin.\cite{mat1985,mat1986} They numerically examined the temporal evolution of a two-dimensional (2D) periodic electric current sheet with imposing the initial fluctuations with a wide range of spectrum. It was reported that the current sheets produce corrugations at large scales and magnetic islands at small scales. Due to these corrugations and magnetic-island motions, multiple X-point structures are generated, which lead to the fast reconnection. It was shown that the fluctuations lead to the fast reconnection through the generation of the multiple X points.

	On the basis of a heuristic argument of the Alfv\'{e}nic turbulence by Goldreich and Sridhar,\cite{gol1995} Lazarian and Vishniac developed a scaling theory of turbulent magnetic reconnection.\cite{laz1999} Due to the stochastic motions of magnetic fields, the effective thickness of the reconnection region becomes much larger, leading to a much larger reconnection rate. Their result suggests that the reconnection rate is not determined by the microphysics quantities such as the molecular diffusivity but by the turbulent quantities. Recently Eyink {\textit{et al.}}\ interpreted the stochastic motions of magnetic fields as the Richardson diffusion, and by using the Lagrangian trajectory formulation, obtained  the reconnection-rate expression equivalent to the Lazarian and Vishniac one.\cite{eyi2011}
	
	Guo {\textit{et al.}}\ examined turbulent reconnection in the framework of the resistive reduced magnetohydrodynamics (RMHD).\cite{guo2012} They adopted a mean-field approach and treat the inhomogeneity of the reconnected magnetic field. They focused on the dynamics of the magnetic helicity density in a state of preexisting turbulence. Using some kinds of closure theory of turbulence, they obtained the expressions for the magnetic reconnection rate in the noisy and Alfv\'{e}n-wave turbulence cases. The reconnection rates is very weakly dependent on the amplitude of the turbulence, which suggests that even with a very weak turbulence, the reconnection can be very fast.

	With the developments of the computer resources, numerical investigations of turbulent magnetic reconnection has been recently developed. Such numerical simulations may be divided into several categories depending on what kind of turbulence is supposed as well as on what kind of geometrical setting is adopted for the simulation. The geometrical settings include the dimensionality (2D or 3D), boundary conditions (periodic or open), etc. On the other hand, the turbulence property includes the following: What is the source of turbulence (instability such as tearing mode, injected one with external forcing, or small-scale reconnection itself)? How does the fluctuation spatially distributed inside and outside the reconnection region? In which scale turbulence is injected or generated?

	Servido {\textit{et al.}}\ investigated turbulent reconnection by numerically solving 2D magnetohydrodynamic (MHD) equations with a periodic boundary condition at high Reynolds number.\cite{ser2009} They found that a large number of X-point structures with various sizes and energies are generated, and a reconnection occurs at each X-point. The reconnection rate ranges from small to large depending on each reconnection event. Statistical analysis of these reconnection events showed that the thickness and width of the reconnection regions correspond to the dissipation and correlation length scales of turbulence, respectively. This result suggests that the magnetic reconnection is subject to the statistical nature of turbulence.

	Kowal {\textit{et al.}}\ performed elaborated 3D numerical simulations of MHD equations with non-periodic boundary conditions and examined turbulence effects on magnetic reconnection.\cite{kow2009,kow2012} They first made the fields evolve without any fluctuations, then at some moment started externally imposing turbulence and saw the temporal evolution of the magnetic reconnection rate. Dependence of the reconnection rate on several factors such as the magnitude of fluctuation, scale of injection, magnitude of molecular diffusivity, guide field, etc.\ was extensively investigated. The basic results were as follows. The magnetic reconnection rate is largely enhanced in the turbulent state, and the rate does not depend on the molecular magnetic diffusivity. It was also shown that the larger the magnitude of fluctuation imposed and the turbulence injection scale is, the larger the reconnection rate becomes. They suggested that several collisionless effects are irrelevant to the reconnection rate in the turbulent state. These results are considered to confirm the idea of fast reconnection due to the magnetic-field wandering proposed by Lazarian and Vishniac.

	Using 2D direct numerical simulations with a periodic boundary condition, Loureiro {\textit{et al.}}\ investigated how the background noise has an effect on the fast reconnection.\cite{lou2009} It was shown that the dependence of the reconnection rate  on the Lundquist number decreases as the turbulence level increases. This result suggests that, in turbulence, the reconnection rate does not depend on the microscopic physics but only on the nonlinear dynamics of turbulence.
	
	Another class of approaches to turbulent reconnection is the plasmoid instability one. A large number of simulation studies have been done to explore the tearing instability of current sheets and to examine the formation of the secondary magnetic island called plasmoid.\cite{dau2009b,uzd2010,nak2013} In this approach, turbulent reconnection is explored without resorting to any pre-existing turbulence. The formation of multiple plasmoid effectively reduces the width of the reconnection region, and consequently contribute to a fast reconnection. Bhattacharjee {\textit{et al.}}\ showed the independence of the reconnection rate on the Lundquist number in the fast reconnection due to the plasmoid instability.\cite{bha2009} Another important point is that the multiple plasmoid generation reduces the length scale to reach a scale smaller than the magnetohydrodynamic limit.\cite{dau2009a} Specific mechanisms based on a hierarchy of plasmoids have been proposed to bridge global scale to dissipation scale in the context of the so-called reconnection phase diagram.\cite{ji2011} How and how much turbulence is generated from laminar current sheets in a fully realistic physical parameter regime is a very important question. How turbulence interacts with the plasmoid instability is still an open question. In the collisionless regime, it is considered that we have not yet seen a significant effect of MHD-scale turbulence on the reconnection rate, nor have we seen evidence of significantly enhanced transport due to MHD turbulence in the numerical simulations where the current sheet scale is of order of ion gyro-radius.\cite{kar2013} At the same time, we should note the fact that there are examples of fast reconnection without a plasmoid. The plasmoid formed in an elongated current sheet can contribute to reduce the inertia resistivity, and the formation of many plasmoids does not necessarily mean the enhancement of magnetic dissipation in a collisionless regime.  It is required to study carefully a macro-micro scale coupling process in a large scale kinetic simulation.  For summary of recent studies and future perspectives, the reader is referred to a review by Karimabadi and Lazarian.\cite{kar2013}
	
	If the turbulence effect is incorporated into the dynamics of the large-scale fields, the behavior of large-scale fields is considerably affected by turbulence through the drastic change of the effective transport. Since turbulence enhances the effective diffusivity, turbulence may enhance the magnetic reconnection very much. However, as was stressed in our previous paper,\cite{yok2011b} this effect is not so self-evident. In most of the previous work that incorporates turbulence effect into the magnetic reconnection process, turbulence is externally imposed or given as a fixed ingredient. This treatment is not sufficient enough. It is true that turbulence determines the dynamics of the mean fields through the effective or turbulence transport. At the same time, the mean-field configurations, which are subject to the turbulence effect, also determine the properties of turbulence. Dynamics of mean-field configuration and turbulence have to be simultaneously treated in a consistent manner.\cite{yok2011b} Considering these points, the obvious questions to be addressed are as follows:

\noindent	
- How and how much turbulence is generated and sustained in the magnetic-reconnection configuration;

\noindent
- How enhanced transport coefficients due to turbulence are spatiotemporally distributed;

\noindent
- How the effective transport is related to the magnetic reconnection rate.

	In order to address these questions, we have to treat the mutual interaction of the mean-field structures and turbulence in a consistent manner. The main objective of this paper is to show a consistent approach to these questions. 

	For this purpose, we adopt the mean-field approach to the turbulent magnetic reconnection in this work since this approach provides a powerful tool for investigating the interplay between the fluctuation and large-scale inhomogeneities in the realistic physical parameter domain. Obviously, the choice of turbulence model is of primary importance. We easily imagine that a simple heuristic turbulence modeling adopting gradient-diffusion approximations such as eddy viscosity, turbulent magnetic diffusivity, etc.\ may not work well. We should remember that a simple eddy-viscosity-type model completely fails even in hydrodynamic case if it is applied to a turbulent swirling flow. Instead, we need turbulence modeling based on the fundamental equations of magnetohydrodynamics. Such models should include turbulence effects other than the simple eddy viscosity and diffusivities. In this work, we adopt a turbulence model on the basis of an elaborated closure theory for inhomogeneous turbulence at very high kinetic and magnetic Reynolds numbers.\cite{yos1990,yok2013} As we see later, the model expressions for the turbulent electromotive force [Eq.~(\ref{eq:emf_exp})] and the Reynolds stress [Eq.~(\ref{eq:rey_stress_exp})] themselves have been confirmed by a direct numerical simulation (DNS) of the Kolmogorov flow with a strong imposed magnetic field.\cite{yok2011c} Also we should note that this type of mean-field or turbulence model approach has been applied to the solar-wind turbulence with a strong mean magnetic field and a fast velocity with shears, and successfully reproduced the evolution of Alfv\'{e}nicity in solar-wind turbulence: the spatial distributions of the turbulent cross helicity and the turbulent MHD residual energy (difference between the kinetic and magnetic energies) observed by satellite.\cite{zho1990,tum1993,yok2007,yok2011a} In this sense, the validity of the mean-field approach has already been confirmed in the context of the magnetic turbulence. The applicability of the mean-field approach in the context of magnetic reconnection will be furhter discussed later.

	The organization of this paper is as follows. In \S~II, we present the fundamental mean-field equations and introduce the most important turbulent correlations appearing in the mean-field equations. In \S~III, the dynamic balance between the enhancement and suppression of transport due to turbulence is explained. In \S~IV, the concept of turbulence model is presented with the explicit expressions for the model equations. In \S~V, features of the mean-field approach to the magnetic reconnection are discussed. In \S~VI, numerical test of the notion of dynamic balance of turbulence is presented in the context of magnetic reconnection. In \S~VII, cross-helicity effects in transport balance are discussed. Conclusions with some suggestions for the future work are presented in \S~VIII.

%-----------------------------------------------------------------------------
%	II Fundamental equations
%-----------------------------------------------------------------------------
\section{Fundamental equations}
%-----------------------------------------------------------------------------
%	II.A Mean equations
%-----------------------------------------------------------------------------
\subsection{Mean equations}
In this work, we treat the magnetic reconnection in the framework of the magnetohydrodynamics (MHD) without resorting to the effects beyond the one-fluid MHD, such as the Hall effect, electron pressure gradient, etc. This treatment does not deny possible importance of such effects. Actually importance of the effects beyond the MHD has been reported by several authors.\cite{bir2007} Here we concentrate our analysis on the MHD with turbulence incorporated and see what consequence we obtain.

	We consider the compressible MHD equations, which consist of the equation of mass, momentum, magnetic-field, and total energy. We divide field quantities into the resolved and the unresolved scales as
\begin{equation}
	f = F + f',\;\; F = \langle f \rangle	
	\label{eq:reynolds_decomp}%(1)
\end{equation}
with $\langle {\cdots} \rangle$ being the ensemble average. Here, the field quantity $f$ denotes
\begin{subequations}
\begin{equation}
	f = \left( {
		{\bf{u}}, \mbox{\boldmath$\omega$}, {\bf{b}}, {\bf{j}}, {\bf{e}}, \rho, p
	} \right)
	\label{eq:inst_fields}%(2a)
\end{equation}
\begin{equation}
	F = \left( {
		{\bf{U}}, \mbox{\boldmath$\Omega$}, {\bf{B}}, {\bf{J}}, {\bf{E}}, 
		\overline{\rho}, P
	} \right)
	\label{eq:mean_fields}%(2b)
\end{equation}
\begin{equation}
	f' = \left( {
	{\bf{u}}', \mbox{\boldmath$\omega$}', {\bf{b}}', {\bf{j}}', {\bf{e}}', \rho', p'
	} \right)
	\label{eq:fluct_fields}%(2c)
\end{equation}
\end{subequations}
where ${\bf{u}}$ is the velocity, $\mbox{\boldmath$\omega$} (= \nabla \times {\bf{u}})$ the vorticity, ${\bf{b}}$ the magnetic field, ${\bf{j}} (= \nabla \times {\bf{b}}/\mu_0)$ the electric-current density, ${\bf{e}}$ the electric field, $\rho$ the density, and $p$ the pressure ($\mu_0$: magnetic permeability).

	Using the Reynolds decomposition [Eq.~(\ref{eq:reynolds_decomp})], the mean-field equations are written as
\begin{equation}
	\frac{\partial \overline{\rho}}{\partial t}
	+ \nabla \cdot \left( {\overline{\rho} {\bf{U}}} \right)
	= - \nabla \cdot \left\langle { \rho' {\bf{u}}'} \right\rangle,
	\label{mean_density_eq}%(3)
\end{equation}
\begin{eqnarray}
	\frac{\partial}{\partial t} \overline{\rho} U^\alpha
	&+& \frac{\partial}{\partial x^a} \overline{\rho} U^a U^\alpha
	\nonumber\\
	\hspace{20pt}&=& - \frac{\partial P}{\partial x^\alpha}
	+ \frac{\partial}{\partial x^\alpha}
	\mu {\cal{S}}^{a \alpha}
	+ \left ( {{\bf{J}} \times {\bf{B}}} \right)^\alpha
	\nonumber\\
	&-& \frac{\partial}{\partial x^a} \left( {
		\overline{\rho} \left\langle {u'{}^a u'{}^\alpha} \right\rangle
		- \frac{1}{\mu_0} \left\langle {b'{}^a b'{}^\alpha} \right\rangle
	}\right.\nonumber\\
	& & \left. { \rule{0.ex}{3.ex}
	+ U^a \left\langle {\rho' u'{}^\alpha} \right\rangle
		+ U^\alpha \left\langle {\rho' u'{}^a} \right\rangle
	} \right)
	+ R_U^\alpha,	
	\label{eq:mean_velocity_eq}%(4)
\end{eqnarray}
\begin{eqnarray}
	&&\frac{\partial}{\partial t} \left( {
	\frac{P}{\gamma_{\rm{s}} - 1}
	+ \frac{1}{2} \overline{\rho} {\bf{U}}^2
	+ \frac{1}{2\mu_0} {\bf{B}}^2
	}\right.\nonumber\\
	&&\hspace{30pt}\left.{
	+ \frac{1}{2} \overline{\rho} \left\langle {
		{\bf{u}}'{}^2 } \right\rangle
	+ \frac{1}{2\mu_0} \left\langle{ {\bf{b}}'{}^2
		}\right\rangle
	+ \langle {
		\rho' {\bf{u}}'} \rangle \cdot {\bf{U}}
		%+ \frac{1}{2}\langle {\rho' {\bf{u}}'{}^2} \rangle
	} \right)
	\nonumber\\
	&&\hspace{0pt}= - \nabla \cdot \left[ {
	\left( {
		\frac{\gamma_{\rm{s}}}{\gamma_{\rm{s}} - 1} P
		+ \frac{1}{2} \overline{\rho} {\bf{U}}^2
		+ \frac{1}{2} \overline{\rho} \langle {{\bf{u}}'{}^2} \rangle
		+ \langle {\rho' {\bf{u}}'} \rangle \cdot {\bf{U}}
		%+ \frac{1}{2} \langle {\rho' {\bf{u}}'{}^2} \rangle
	} \right) {\bf{U}}
	} \right.\nonumber\\
	&&
	\left. {
	\hspace{40pt} + \left\langle {
	\left( {
		\frac{\gamma_{\rm{s}}}{\gamma_{\rm{s}} - 1} p'
		+ \overline{\rho}{\bf{U}}\cdot {\bf{u}}'
		%+ \frac{1}{2} \overline{\rho} {\bf{u}}'{}^2
		+ \frac{1}{2} \rho' {\bf{U}}^2
		%+ \rho' {\bf{u}}' \cdot {\bf{U}}
		%+ \frac{1}{2} \rho' {\bf{u}}'{}^2
	} \right) {\bf{u}}'
	} \right\rangle
	} \right.\nonumber\\
	&&\left. {
		\hspace{40pt}
		+ \frac{{\bf{E}} \times {\bf{B}}}{\mu_0}
		%+ \frac{\langle {{\bf{e}}' \times {\bf{b}}'} \rangle}{\mu_0}
	} \right]
	+ R_E,	
	\label{eq:mean_energy_eq}%(5)
\end{eqnarray}
\begin{equation}
	\frac{\partial {\bf{B}}}{\partial t}
	= - \nabla \times {\bf{E}},	
	\label{eq:mean_faraday_eq}%(6)
\end{equation}
\begin{equation}
	{\bf{E}}
	= - {\bf{U}} \times {\bf{B}}
	+ \eta {\bf{J}}
	- {\bf{E}}_{\rm{M}},	
	\label{eq:mean_ohms_law}%(7)
\end{equation}
where $\mu$ is the dynamic viscosity, $\gamma_{\rm{s}} (= C_P / C_V)$ the ratio of the specific heats ($C_P$: specific heat at constant pressure, $C_V$: the specific heat at constant volume) and  ${\mbox{\boldmath${\cal{S}}$}}$ is the strain rate of the mean velocity defined by
\begin{equation}
	{\cal{S}}^{\alpha\beta}
	= \frac{\partial U^\beta}{\partial x^\alpha}
	+ \frac{\partial U^\alpha}{\partial x^\beta}
	- \frac{2}{3}\nabla\cdot{\bf{U}}\delta^{\alpha\beta}.
	\label{eq:mean_vel_strain}%(8)
\end{equation}
In Eqs.~(\ref{eq:mean_velocity_eq}) and (\ref{eq:mean_energy_eq}), ${\bf{R}}_{U}$ and $R_{E}$ are defined by
\begin{equation}
	R_U^\alpha 
	= - \frac{\partial}{\partial t}\langle {\rho' u'{}^\alpha} \rangle
	- \frac{\partial}{\partial x^a} \langle {\rho' u'{}^a u'{}^\alpha} \rangle
	- \frac{1}{2 \mu_0} \frac{\partial}{\partial x^\alpha} \langle {{\bf{b}}'{}^2} \rangle,
	\label{eq:residual_U}%(9)
\end{equation}
\begin{eqnarray}
	R_E &=& - \frac{\partial}{\partial t} \langle {\rho' {\bf{u}}'{}^2} \rangle
		- \nabla \cdot \left( {
		\frac{1}{2} \langle {\rho' {\bf{u}}'{}^2} \rangle {\bf{U}}	
		+ \frac{1}{2} \overline{\rho} \langle {{\bf{u}}'{}^2 {\bf{u}}'} \rangle
	} \right.
	\nonumber\\
	&&\hspace{0pt}\left. {
		+ \langle {\rho' {\bf{u}}' \cdot {\bf{U}} {\bf{u}}'} \rangle
		+ \frac{1}{2} \langle {\rho' {\bf{u}}'{}^2 {\bf{u}}'} \rangle
		+ \frac{\langle {{\bf{e}}' \times {\bf{b}}'} \rangle}{\mu_0}
	} \right).
	\label{eq:residual_E}%(10)
\end{eqnarray}
They consist of the terms that may be neglected compared with the retained ones. In Eq.~(\ref{eq:mean_ohms_law}), ${\bf{E}}_{\rm{M}}$ is the turbulent electromotive force defined by
\begin{equation}
	{\bf{E}}_{\rm{M}}
	\equiv \left\langle {{\bf{u}}' \times {\bf{b}}'} \right\rangle.
	\label{eq:emf_def}%(11)
\end{equation}
Note that substitution of Eq.~(\ref{eq:mean_ohms_law}) into Eq.~(\ref{eq:mean_faraday_eq}) leads to the mean magnetic-field induction equation:
\begin{equation}
	\frac{\partial {\bf{B}}}{\partial t}
	= \nabla \times \left( {
		{\bf{U}} \times {\bf{B}} + {\bf{E}}_{\rm{M}}
	} \right)
	+ \eta \nabla^2 {\bf{B}}.
	\label{eq:mean_ind_eq}%(12)
\end{equation}

	Equations~(\ref{mean_density_eq})-(\ref{eq:mean_ind_eq}) show that the correlations of fluctuations such as $\langle {\bf{u}}' {\bf{u}}' \rangle$, $\langle {\bf{b}}' {\bf{b}}' \rangle$, $\langle {\rho' {\bf{u}}'} \rangle$, $\langle p' {\bf{u}}' \rangle$,  etc.\ contribute to the transport of the mean fields.

%-----------------------------------------------------------------------------
%	II.B Reynolds stress and turbulent electromotive force
%-----------------------------------------------------------------------------
\subsection{Reynolds stress and turbulent electromotive force}
If we neglect the fluctuation of the density and pressure as
\begin{subequations}\label{eq:no_fluct_rho_p}%(13)
\begin{equation}
	\rho = \overline{\rho} + \rho' = \overline{\rho},\;\; \rho' = 0,
	\label{eq:no_fluct_density}%(13a)
\end{equation}
\begin{equation}
	p = P + p' = P,\;\; p' = 0,
	\label{eq:no_fluct_pressure}%(13b)
\end{equation}
\end{subequations}
then the relevant correlations are only the turbulent electromotive force ${\bf{E}}_{\rm{M}}$ [Eq.~(\ref{eq:emf_def})] and the Reynolds stress $\mbox{\boldmath${\cal{R}}$}$ defined by
\begin{equation}
	{\cal{R}}^{\alpha\beta} 
	\equiv \left\langle {
	u'{}^\alpha u'^\beta 
	- \frac{1}{\mu_0 \overline{\rho}} b'^\alpha b'^\beta
	} \right\rangle.
	\label{eq:rey_stress_def}%(14)
\end{equation}
This treatment does not deny the importance of the fluctuations in density and pressure (or temperature, internal energy). Actually in several situations, the turbulent mass and heat transports $\langle {\rho' {\bf{u}}'} \rangle$ and $\langle {p' {\bf{u}}'} \rangle$ (or $\langle {\theta' {\bf{u}}'} \rangle$, $\langle {q' {\bf{u}}'} \rangle$) play important roles in the mean-field transportation ($\theta'$: temperature fluctuation, $q'$: internal energy fluctuation). The present treatment is just a first step of self-consistent turbulence approach to the magnetic reconnection.

	For the sake of simplicity, hereafter we rewrite the magnetic field, electric-current density, etc.\ using the Alfv\'{e}n-speed unit as
\begin{equation}
	\frac{\bf{b}}{(\mu_0 \overline{\rho})^{1/2}}
	\to {\bf{b}},\;\;
	\frac{\bf{j}}{(\overline{\rho}/\mu_0)^{1/2}}
	\to {\bf{j}},\;\;
	\frac{\bf{e}}{(\mu_0 \overline{\rho})^{1/2}}
	\to {\bf{e}},\;\;
	\frac{p}{\overline{\rho}}
	\to p.
	\label{eq:alfven_units}%(15)
\end{equation}
 Under condition (\ref{eq:no_fluct_rho_p}), the expressions for $\mbox{\boldmath${\cal{R}}$}$ and ${\bf{E}}_{\rm{M}}$ become the same as for the counterparts in the incompressible turbulence case. They are expressed as\cite{yos1990,yok2013}
 \begin{equation}
	{\bf{E}}_{\rm{M}}
	=  - \beta {\bf{J}} + \gamma \mbox{\boldmath$\Omega$} + \alpha {\bf{B}},
	\label{eq:emf_exp}%(16)
\end{equation}
\begin{eqnarray}
	&&\left[ {{\cal{R}}^{\alpha\beta}} \right]_{\rm{D}}
	\equiv {\cal{R}}^{\alpha\beta} 
	- \frac{2}{3} K_{\rm{R}} \delta^{\alpha\beta}
	\nonumber\\
	&&\hspace{10pt}= - \nu_{\rm{K}} {\cal{S}}^{\alpha\beta}
	+ \nu_{\rm{M}} {\cal{M}}^{\alpha\beta}
	+ \left[ {
		\Gamma^\alpha \Omega^\beta 
		+ \Gamma^\beta \Omega^\alpha
	} \right]_{\rm{D}},
	\label{eq:rey_stress_exp}%(17)
\end{eqnarray}
where $[{\cal{A}}^{\alpha\beta}]_{\rm{D}}$ is the deviatoric part of ${\cal{A}}^{\alpha\beta}$, $[{\cal{A}}^{\alpha\beta}]_{\rm{D}} = {\cal{A}}^{\alpha\beta} - {\cal{A}}^{aa} \delta^{\alpha\beta}/3$, and $K_{\rm{R}} = \langle {{\bf{u}}'{}^2 - {\bf{b}}'{}^2} \rangle/2$ is the turbulent MHD residual energy. In Eqs.~(\ref{eq:emf_exp}) and (\ref{eq:rey_stress_exp}), $\mbox{\boldmath$\Omega$} (= \nabla \times {\bf{U}})$ is the mean vorticity and $\mbox{\boldmath${\cal{M}}$}$ is the strain rate of the mean magnetic field defined by
\begin{equation}
	{\cal{M}}^{\alpha\beta}
	= \frac{\partial B^\beta}{\partial x^\alpha}
	+ \frac{\partial B^\alpha}{\partial x^\beta}
	- \frac{2}{3} \nabla\cdot{\bf{B}}\delta^{\alpha\beta}
	\label{eq:mean_mag_strain}%(18)
\end{equation}
(note that we adopted the Alfv\'{e}n-speed unit and $\nabla \cdot {\bf{B}}$ does not necessarily vanish).

	The transport coefficients $\nu_{\rm{K}}$ and $\beta$ are related to the turbulent MHD energy $K (\equiv \langle {{\bf{u}}'{}^2 + {\bf{b}}'{}^2} \rangle /2)$, $\nu_{\rm{M}}$ and $\gamma$ are related to the turbulent cross helicity $W (\equiv \langle {{\bf{u}}' \cdot {\bf{b}}'} \rangle)$, and $\alpha$ and $\mbox{\boldmath$\Gamma$}$ are related to the turbulent residual helicity  $H (\equiv \langle {- {\bf{u}}' \cdot \mbox{\boldmath$\omega$}' + {\bf{b}}' \cdot {\bf{j}}'} \rangle)$ and its inhomogeneity $\nabla H$, respectively. They are related to each other and modeled as
\begin{subequations}\label{eq:trans_coeff_rel}%(19)
\begin{equation}
	\beta = \frac{5}{7} \nu_{\rm{K}} = C_\beta \tau K,
	\label{eq:beta_nuk_rel}%(19a)
\end{equation}
\begin{equation}
	\gamma = \frac{5}{7} \nu_{\rm{M}} = C_\gamma \tau W,
	\label{eq:gamma_num_rel}%(19b)
\end{equation}
\begin{equation}
	\alpha = C_\alpha \tau H,
	\label{eq:alpha_rel}%(19c)
\end{equation}
\end{subequations}
where $\tau$ is the timescale of turbulence and $C_\beta$, $C_\gamma$, and $C_\alpha$ are model constants of the same order of magnitude each other, $O(10^{-1})$. In this work, we drop $\alpha$ and $\mbox{\boldmath$\Gamma$}$-related terms since there is no helicity generation in the present situation. This point will be referred to later in \S~VI.

	In the model expressions [Eqs.~(\ref{eq:emf_exp}) and (\ref{eq:rey_stress_exp})], the transport coefficients are scalars. This does not deny the importance of anisotropy in magnetic turbulence. As some heuristic models showed,\cite{gol1995} anisotropy is considered to play a very important role in MHD turbulence evolution especially in the context of the turbulent magnetic reconnection.\cite{laz1999,eyi2011} 

	However, we should note the following points. Firstly, even by using isotropic transport coefficients, we can well express the effects of anisotropy. This is because we solve the evolution equations for the transport coefficients (or turbulent quantities that express the transport coefficients). Those equations contain several mean fields and their inhomogeneities. In the present work, the transport coefficients are determined by the inhomogeneities and anisotropy of the mean fields such as the magnetic field and velocity shears through the transport equations of the turbulent statistical quantities. For example, in the Alfv\'{e}nic turbulence fluctuations are constituted by the collection of Alfv\'{e}n waves. Such a property is intrinsically incorporated into the transport equations of turbulence quantities, which are derived from the full MHD equations. Asymmetry of the Alfv\'{e}n waves propagating parallel and antiparallel to the mean magnetic field leads to a finite turbulent cross helicity. This effect is implemented into the present turbulence model through the term ${\bf{B}} \cdot \nabla K$ (energy inhomogeneity along the magnetic field) in Eq.~(\ref{eq:T_W_def}). 
	
	Secondly, as in the magnetic turbulence case, it is often assumed in the theoretical and numerical studies of planetary atmosphere that the viscosity is anisotropic between the horizontal and vertical directions. However, it is also shown that the isotropic eddy-viscosity formulation associated with the transport equations of turbulence quantities and timescales can reproduce the planetary superrotation without resorting any anisotropic viscosities.\cite{yos2013} This fact clearly shows that an isotropic formulation of the turbulent viscosity properly implemented with the mean-field inhomogeneity arguments can treat anisotropic transport phenomena. In this sense, our present model is not for homogeneous and isotropic turbulence but for inhomogeneous and anisotropic turbulence.

%-----------------------------------------------------------------------------
%	III Transport enhancement and suppression due to turbulence
%-----------------------------------------------------------------------------
\section{Transport enhancement and suppression due to turbulence}
	The structures of the expressions for ${\bf{E}}_{\rm{M}}$ [Eq.~(\ref{eq:emf_exp})] and $\mbox{\boldmath${\cal{R}}$}$ [Eq.~(\ref{eq:rey_stress_exp})] are similar each other. The first or energy-related terms ($\beta$ and $\nu_K$) enhance the transport. On the other hand, the second or cross-helicity-related terms ($\gamma$ and $\nu_{\rm{M}}$) as well as the third or helicity-related terms ($\alpha$, $\mbox{\boldmath$\Gamma$}$) may suppress the transport. For example, if we substitute Eq.~(\ref{eq:emf_exp}) into Eq.~(\ref{eq:mean_ind_eq}), we obtain
\begin{eqnarray}
	\frac{\partial {\bf{B}}}{\partial t}
	&=& \nabla \times \left( {{\bf{U}} \times {\bf{B}}} \right)
	- \nabla \times \left[ {
		\left( {\eta + \beta} \right) \nabla \times {\bf{B}}
	} \right]\nonumber\\
	&& + \nabla \times \left( {
		\gamma \mbox{\boldmath$\Omega$} + \alpha {\bf{B}}
	} \right).
	\label{eq:mean_ind_exp}%(20)
\end{eqnarray}
As can be seen from $\eta \to \eta + \beta$ in Eq.~(\ref{eq:mean_ind_exp}), the $\beta$-related term in Eq.~(\ref{eq:emf_exp}) represents the transport enhancement due to turbulence, which is called the turbulent magnetic diffusivity. As Eq.~(\ref{eq:beta_nuk_rel}) shows, in the presence of turbulence ($K \ne 0$), we always have the turbulent magnetic diffusivity $\beta$ (and the turbulent viscosity $\nu_{\rm{K}}$) which is much larger than the molecular counterpart $\eta$ (and $\nu$).

	At the same time, however, we may have some effects of pseudoscalars arising from the breakage of symmetry in turbulence. The second and third or $\gamma$- and $\alpha$-related terms in Eq.~(\ref{eq:mean_ind_exp}) may contribute to the magnetic-field generation. If we have a large cross correlation between the velocity and magnetic field in turbulence, it may act for balancing the enhanced magnetic-diffusivity (and turbulent viscosity) effect [Eqs.~(\ref{eq:emf_exp}) and (\ref{eq:rey_stress_exp})]. If the $\gamma$- and $\alpha$-related terms are nearly balanced with the $\beta$-related term, we have Eq.~(\ref{eq:mean_ind_eq}) with ${\bf{E}}_{\rm{M}} \simeq 0$ as
\begin{equation}
	\frac{\partial {\bf{B}}}{\partial t}
	\simeq \nabla \times \left( {
		{\bf{U}} \times {\bf{B}}
	} \right)
	+ \eta \nabla^2 {\bf{B}}.
	\label{eq:mean_ind_eq_wo_emf}%(21)
\end{equation}
This is the same equation in form as the laminar induction equation. This suggests that, if the magnetic Reynolds number is huge ($Rm \gg 1$), the mean magnetic flux can be frozen in the fluid motion even in a highly turbulent state. This is a typical situation of the dynamic balance between the transport enhancement and suppression due to turbulence. We should note that although the equation and consequently the magnetic field behaviors appear very similar to the counterparts in the laminar case, the underlying physics of Eq.~(\ref{eq:mean_ind_eq_wo_emf}) is entirely different from that of the laminar induction equation. In the dynamic balance case, where turbulence itself is present, once the balance between the enhancement and suppression is broken, we usually encounter a strong transport enhancement depending on the level of turbulence.

	The turbulence effect including the enhanced transport results from the properties of flow but not of fluid. So the transport coefficients in general depend on space and time. If we have no or very small pseudoscalar quantities in space or in time, then the enhancement of transport becomes dominant. There the turbulent electromotive force is expected to be represented only by the turbulent magnetic diffusivity as
\begin{equation}
	{\bf{E}}_{\rm{M}} \simeq - \beta \nabla \times {\bf{B}}.
	\label{eq:no_dynamo_emf}%(22)
\end{equation}
Then the induction equation for the mean magnetic field is written as
\begin{equation}
	\frac{\partial {\bf{B}}}{\partial t}
	\simeq \nabla \times \left( {
		{\bf{U}} \times {\bf{B}}
	} \right)
	- \nabla \times \left( {
		\beta \nabla \times {\bf{B}}
	} \right),
	\label{eq:no_dynamo_mean_ind_eq}%(23)
\end{equation}
where the molecular magnetic diffusivity $\eta$ has been dropped since $\eta \ll \beta$. Only in such a region, we can expect the dominant effect of the turbulent magnetic diffusivity.

	As for more fundamental discussions about the breakage condition of the magnetic-flux freezing, the reader is referred to a recent paper by Eyink {\textit{et al.}}\ (2011), where the flux-freezing of general magnetic fields (not mean fields) is well discussed from the viewpoint of the spontaneous stochasticity.\cite{eyi2011}

	Dynamic balance and non-balance between the transport enhancement and suppression is one of the most interesting features of turbulence. We see that the transport property due to turbulence is determined by the spatiotemporal evolution of the turbulent statistical quantities such as the turbulent MHD energy, cross helicity, and residual helicity. At the same time, as was stressed in \S~I and Ref.~\onlinecite{yok2013}, the evolution of turbulence is determined by the large-scale or mean-field configurations. Dynamics of the mean fields and turbulence should be solved simultaneously. As for the other arguments related to the mitigation of turbulence mixing through the change of cascading rates due to the cross helicity, the reader is referred to the analysis of the imbalanced Alfv\'{e}nic turbulence.\cite{mar2001,cho2002,ber2008}

\begin{figure}[htb]
\includegraphics[width=.35\textwidth]{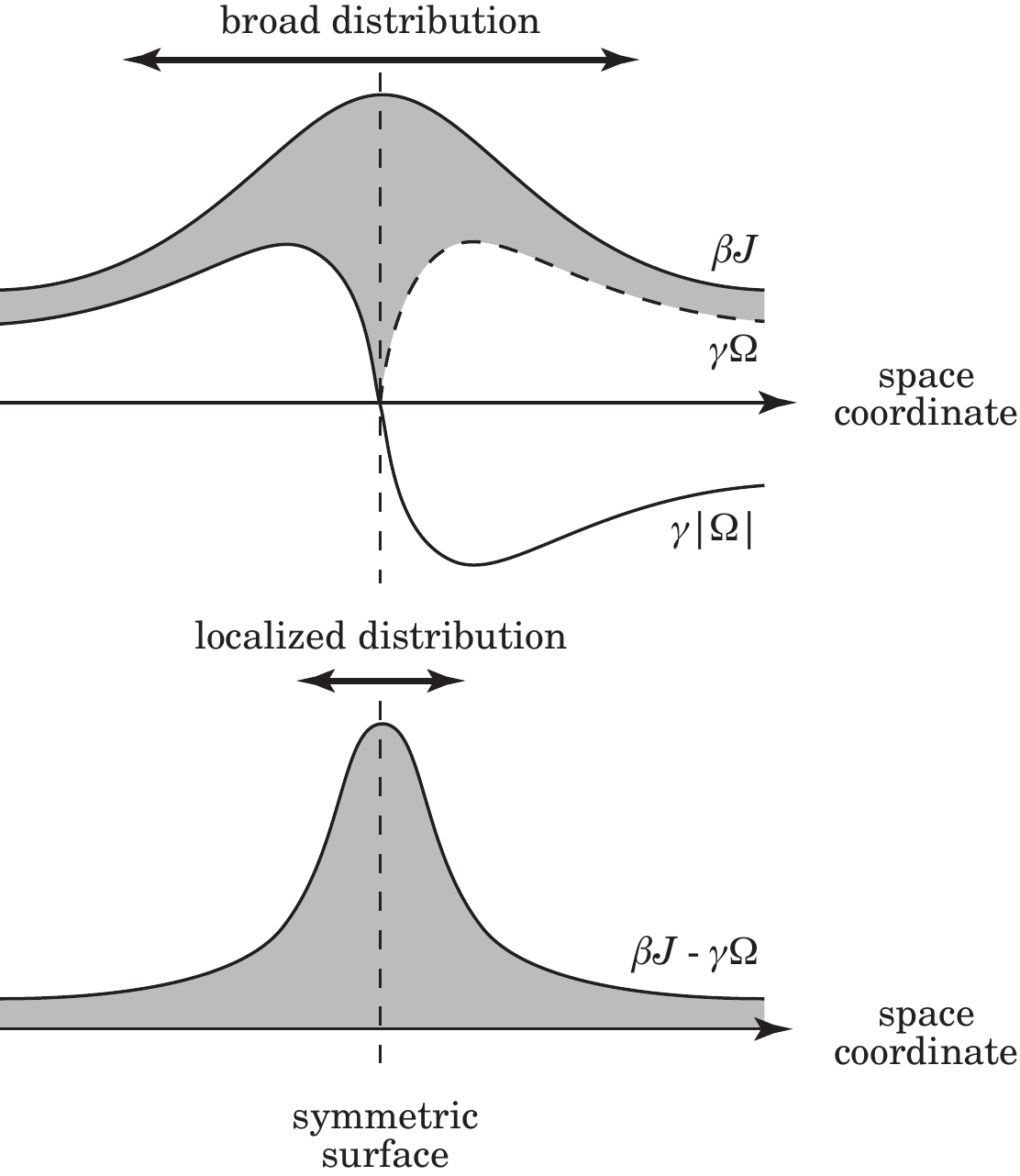}
\caption{\label{fig:dyn_balance} Schematic spatial distributions of the $\beta$- and $\gamma$-related terms. Dynamic balance between the transport enhancement due to turbulent energy $\beta {\bf{J}}$ and the suppression due to the turbulent cross helicity $\gamma \mbox{\boldmath$\Omega$}$.}
\end{figure}

	As was predicted by our previous paper,\cite{yok2011b} even if the total amount or volume average of the turbulent cross helicity is zero, the turbulent cross helicity can be spatially or locally distributed positive and negative depending on the mean-field configuration. Since it is a consequence of the symmetric and antisymmetric properties of the velocity and the magnetic field, the quadrupole-like spatial distribution of the cross helicity is ubiquitous around the current sheet.\cite{men1996} This is a prominent feature of pseudoscalar quantities, which is non-positive definite unlike pure scalars such as the energy. If the turbulent cross helicity changes its sign on the symmetry surface, its magnitude must be very small in the vicinity of the surface. This means that the transport suppression due to the turbulent cross helicity can not function in the vicinity of the symmetry surface, hence only the transport enhancement due to turbulence can effectively function there. Again, this is a consequence of the dynamic (non)balance between the transport enhancement due to the turbulent energy and the transport suppression due to the turbulent cross helicity. The dynamic balance of turbulence may contribute to the strong localization of the effective turbulent magnetic diffusivity, leading to the fast magnetic reconnection. Schematic situation is depicted in Fig.~\ref{fig:dyn_balance}. Even if the spatial distribution of $\beta$ is broad, the balancing with $\gamma$ may lead to a fairly localized distribution of the turbulent magnetic diffusivity.

%------------------------------------------
%	Fig. 1
%------------------------------------------

	The notion of dynamic balance between the enhancement and suppression of transport should be compared with the notion of the dynamic aligment of the magnetohydrodynamic turbulence.\cite{bol2006,mas2008} In the latter, basically a relaxation problem of the magnetohydrodynamic turbulence with a finite flow under the constraint of the total amount of the cross helicity is argued. It is numerically established that the angle between the velocity and magnetic-field fluctuations decreases as the scale decreases (dynamic alignment). Numerical simulations are designed in the framework of the homogeneous turbulence without any inhomogeneous mean flow and the turbulence is sustained by the external forcing. On the other hand, in the dynamic balance argument, we treat inhomogeneous turbulence with mean velocity and magnetic-field inhomogeneities. These inhomogeneities in the mean fields are the main sources of the turbulent MHD energy and the turbulent cross helicity.
	
	We should also remark the following points.\cite{yok2013} 
	
	First, the relationship between the magnitude of the turbulent EMF $\langle {{\bf{u}}' \times {\bf{b}}'} \rangle$ and the turbulent cross helicity $\langle {{\bf{u}}' \cdot {\bf{b}}' } \rangle$ is not so simple. The turbulent electromotive force (EMF) $\langle {{\bf{u}}' \times {\bf{b}}'} \rangle$ [Eq.~(\ref{eq:emf_def})] is the most important quantity representing the turbulence effects on the mean magnetic-field evolution. In relation to the turbulent cross helicity $\langle {{\bf{u}}' \cdot {\bf{b}}'} \rangle$, it is often argued that the cross helicity should be small in situations where the turbulence plays an important role in transport through the EMF. This argument is simply based on the fact that the turbulent cross helicity represents the cosine of ${\bf{u}}'$ and ${\bf{b}}'$ while the turbulent EMF does sine of them:
\begin{equation}
	\frac{({\bf{u}}' \cdot {\bf{b}}')^2}{|{\bf{u}}'|^2 |{\bf{b}}'|^2} 
	+ \frac{({\bf{u}}' \times {\bf{b}}')^2}{|{\bf{u}}'|^2 |{\bf{b}}'|^2} 
	= 1.
	\label{eq:cos_sin}%(24)
\end{equation}
However, the situation in general is not so simple since both EMF and turbulent cross helicity are statistically averaged quantities and the cross helicity is not a positive-definite quantity. This point can be easily seen if we consider a case where the turbulent velocity and magnetic fields are totally aligned but the numbers of positive and negative alignments are completely equivalent. In this ``balanced'' turbulence case, the turbulent EMF vanishes ($\langle {{\bf{u}}' \times {\bf{b}}'} \rangle = 0$) due to the alignment. At the same time, the cross helicity vanishes ($\langle {{\bf{u}}' \cdot {\bf{b}}'} \rangle = 0$) due to the equivalence between the positive and negative alignments. In this specific case, both the turbulent EMF and the turbulent cross helicity vanish simultaneously. This example shows that the relationship between the turbulent EMF and the turbulent cross helicity is not so simple.

	Secondly, we do not require a large magnitude of the turbulent EMF for the dynamic balance situation. The magnitude of the turbulent EMF is determined by the balance between the turbulent magnetic diffusivity term, $\beta {\bf{J}}$, and the pseudoscalar-related terms, $\gamma \mbox{\boldmath$\Omega$}$ and $\alpha {\bf{B}}$. If the turbulent cross helicity is large, the turbulent EMF may be small or large depending on the balance with the $\beta {\bf{J}}$ term.
	
	Thirdly, the normalized turbulent cross helicity $\langle {{\bf{u}}' \cdot {\bf{b}}'} \rangle / (\sqrt{{\bf{u}}'{}^2} \sqrt{{\bf{b}}'{}^2})$ represents the degree of alignment between the velocity and magnetic-field fluctuations. In addition, another normalized cross helicity defined by $2\langle {{\bf{u}}' \cdot {\bf{b}}'} \rangle / \langle {{\bf{u}}'{}^2 + {\bf{b}}'{}^2} \rangle$ is of crucial importance if we consider the transport property of turbulence [see Eq.~(\ref{eq:scaled_cross_hel})]. The former is a measure of the topological property while the latter is the counterpart of the dynamical one. Actually in the case of dynamic balance between the transport enhancement and suppression, the key quantity is not the alignment angle $\langle {{\bf{u}}' \cdot {\bf{b}}'} \rangle / (\sqrt{{\bf{u}}'{}^2} \sqrt{{\bf{b}}'{}^2})$ but the relative cross helicity defined by $2\langle {{\bf{u}}' \cdot {\bf{b}}'} \rangle / \langle {{\bf{u}}'{}^2 + {\bf{b}}'{}^2} \rangle$.

%-----------------------------------------------------------------------------
%	IV Turbulence model
%-----------------------------------------------------------------------------
\section{Turbulence model}
%-----------------------------------------------------------------------------
%	IV.A Basic concept
%-----------------------------------------------------------------------------
\subsection{Basic concept}
The most straightforward approach to turbulence is direct numerical simulations (DNSs), where fundamental equations are directly solved without resorting any types of model. In order to resolve inhomogeneous and anisotropic turbulence with realistic boundary conditions, we have to use a huge number of grid points. The ratio of the largest scale in turbulence motions ($\ell_{\rm{E}}$: energy-containing scale) to the smallest ones ($\ell_{\rm{D}}$: Kolmogorov scale) is scaled by the Reynolds number $Re$ as $\ell_{\rm{E}} / \ell_{\rm{D}} \sim Re^{3/4}$. This means that we need the number of grid points of $O(Re^{9/4})$ to resolve all scales of turbulence motion. If the Reynolds number is huge as in astro- and space physics phenomena, we need a huge number of grid points. This is not available in the foreseeable future even with the present progress of the computer resources.

	Turbulence modeling provides a powerful tool for treating realistic turbulence with mean-field inhomogeneities. Depending on how much turbulent motion is resolved in the simulation, turbulence models are divided into a few categories.
	
	If we solve large scales of turbulent motions with low-pass filter and modeling for small-scale motions, such kind of simulations and modeling are called the large-eddy simulation (LES) and the sub-grid scale (SGS) modeling, respectively. In the LES only the large eddies or large-scale turbulent motions are resolved. The resolved motion is called the grid scale (GS) one, and the small-scale or SGS motions of turbulence should be modeled. Small-scale turbulent motions are expected to have more universal and symmetric properties than large-scale ones. As this result, the SGS motions can be modeled in a form simpler than the case of the Reynolds-averaged turbulence model, where all scales of turbulent motions should be modeled. The simplest SGS model is the Smagorinsky model, where, in addition to the filter width $\Delta$, only the GS velocity strain rate is used for modeling of the SGS viscosity. Note that the filter width $\Delta$ represents the largest scale of the unresolved motions.
	
	Another type of model is called the Reynolds-averaged turbulence model. In this model, field quantities are divided into the mean and the deviation from it. In the simulation, only the mean-field quantities are solved with turbulent transport coefficients such as the eddy or turbulent viscosity, turbulent diffusivity, etc. These transport coefficients represents statistical properties of fluctuation fields, whose motions are not resolved. Unlike the large-eddy simulations (LESs), in the Reynolds-averaged model, all scales of turbulent motions are modeled using a few statistical quantities such as the mixing length, the turbulent energy, its dissipation rate, etc. These quantities are expected to represent statistical properties of turbulence motions.
	
	In order to close the system of model equations, the transport coefficients appearing in the mean-field equations should be expressed in terms of the turbulent statistical quantities. The simplest model is the mixing-length theory, where the turbulent viscosity is modeled as $\nu_{\rm{T}} \sim u \ell \sim \ell^2 |\nabla {\bf{U}}|$, using the mixing length $\ell$ ($u$: characteristic turbulence velocity, ${\bf{U}}$: mean velocity). Depending on the phenomena, the length scale characteristic to the particular situation is adopted as the mixing length. Typical choices are the dimension of the object (for flows around a body), the pressure or density scale hight (for the solar convection zone), the disk thickness (for accretion disks), and so on.

%-----------------------------------------------------------------------------
%	IV.B Choice of turbulence statistical quantities
%-----------------------------------------------------------------------------
\subsection{Choice of turbulence statistical quantities}
The transport coefficients appearing in the mean-field equations should reflect statistical properties of turbulence. In order to construct a self-consistent turbulence model, we have to properly treat the dynamics of turbulence simultaneously with those of the mean fields. In order to represent the turbulence properties, we need quantities that properly represent turbulence. In this sense, choice of such quantities is of essential importance in turbulence modeling.

	As was referred to in \S~II.B, the transport coefficients $\beta$, $\gamma$, and $\alpha$ are related to the turbulent MHD energy, the turbulent cross helicity, and the turbulent residual helicity, respectively [Eq.~(\ref{eq:trans_coeff_rel})]. So, it is natural to adopt them as the turbulence statistical quantities. From the viewpoint of turbulence modeling, relevant quantities are ones that characterize the level of fluctuations. We define
\begin{equation}
	K = \frac{1}{2} \left\langle {{\bf{u}}'{}^2 + {\bf{b}}'{}^2} \right\rangle,
	\label{eq:K_def}%(25)
\end{equation}
\begin{equation}
	W = \left\langle {{\bf{u}}' \cdot {\bf{b}}'} \right\rangle,	
	\label{eq:W_def}%(26)
\end{equation}
\begin{equation}
	H = \left\langle {
	- {\bf{u}}' \cdot \mbox{\boldmath$\omega$}'
	+ {\bf{b}}' \cdot {\bf{j}}'
	} \right\rangle,
	\label{eq:H_def}%(27)
\end{equation}
and denote them as the turbulent MHD energy, turbulent cross helicity, and the turbulent residual helicity. As mentioned above, we do not treat the $\alpha$- or $H$-related effect in this work. The model equation for $H$ and its simplified version have been proposed.\cite{yos1996,yok2008}

	Note that conceptionally and numerically we do not presume any turbulence quantities such as $K$, $W$, and $H$ from the beginning. These turbulence quantities evolve according to the transport equations of them. If there is no generation mechanisms associated with large-scale inhomogeneities [typically written as Eqs.~(\ref{eq:P_K_exp}) for $K$ and (\ref{eq:P_W_exp}) for $W$], then we just do not have that quantity in turbulence.

	From the equations for the velocity and magnetic-field fluctuations, we obtain the evolution equations of $K$ and $W$ as
\begin{equation}
	\frac{DG}{Dt}
	\equiv \left( {
	\frac{\partial}{\partial t} + {\bf{U}} \cdot \nabla 
	} \right) G
	= P_G - \varepsilon_G + T_G,
	\label{eq:K-W_eq}%(28)
\end{equation}
with $G=(K,W)$. Here $P_G$, $\varepsilon_G$, and $T_G$ are the production, dissipation, and transport rates of the quantity $G$. They are defined by
\begin{subequations}\label{eq:K_eq_terms}%(29)
\begin{equation}
	P_K
	= - {\bf{E}}_{\rm{M}} \cdot {\bf{J}}
	- {\cal{R}}^{ab} \frac{\partial U^a}{\partial x^b},
	\label{eq:P_K_def}%(29a)
\end{equation}
\begin{equation}
	\varepsilon_K
	= \nu \left\langle {
		\frac{\partial u'{}^a}{\partial x^b}
		\frac{\partial u'{}^a}{\partial x^b}
	} \right\rangle
	+ \eta \left\langle {
		\frac{\partial b'{}^a}{\partial x^b}
		\frac{\partial b'{}^a}{\partial x^b}
	} \right\rangle
	(\equiv \varepsilon),
	\label{eq:eps_K_def}%(29b)
\end{equation}
\begin{equation}
	T_K
	= {\bf{B}} \cdot \nabla W
	+ \nabla \cdot {\bf{T}}'_K,
	\label{eq:T_K_def}%(29c)
\end{equation}
\end{subequations}
\begin{subequations}\label{eq:W_eq_terms}%(30)
\begin{equation}
	P_W
	= - {\bf{E}}_{\rm{M}} \cdot \mbox{\boldmath$\Omega$}
	- {\cal{R}}^{ab} \frac{\partial B^a}{\partial x^b},
	\label{eq:P_W_def}%(30a)
\end{equation}
\begin{equation}
	\varepsilon_W
	= (\nu + \eta) \left\langle {
		\frac{\partial u'{}^a}{\partial x^b}
		\frac{\partial b'{}^a}{\partial x^b}
	} \right\rangle,
	\label{eq:eps_W_def}%(30b)
\end{equation}
\begin{equation}
	T_W
	= {\bf{B}} \cdot \nabla K
	+ \nabla \cdot {\bf{T}}'_W.	
	\label{eq:T_W_def}%(30c)
\end{equation}
\end{subequations}
The production rates, $P_K$ and $P_W$, arise from the coupling of the turbulence correlations with the mean-field inhomogeneities. If we substitute the expressions for ${\bf{E}}_{\rm{M}}$ [Eq.~(\ref{eq:emf_exp})] and $\mbox{\boldmath${\cal{R}}$}$ [Eq.~(\ref{eq:rey_stress_exp})] into Eqs.~(\ref{eq:P_K_def}) and (\ref{eq:P_W_def}), we obtain the concrete expressions for the production rates as
\begin{equation}
	P_K
	= \beta {\bf{J}}^2
	- \gamma \mbox{\boldmath$\Omega$} \cdot {\bf{J}}
	+ \frac{1}{2} \nu_{\rm{K}} \mbox{\boldmath${\cal{S}}$}^2
	- \frac{1}{2} \nu_{\rm{M}} \mbox{\boldmath${\cal{S}}$}
		:\mbox{\boldmath${\cal{M}}$},
	\label{eq:P_K_exp}%(31)
\end{equation}
\begin{equation}
	P_W
	= \beta \mbox{\boldmath$\Omega$} \cdot {\bf{J}}
	- \gamma \mbox{\boldmath$\Omega$}^2
	+ \frac{1}{2} \nu_{\rm{K}} \mbox{\boldmath${\cal{S}}$}
		:\mbox{\boldmath${\cal{M}}$}
	- \frac{1}{2} \nu_{\rm{M}} \mbox{\boldmath${\cal{M}}$}^2.
	\label{eq:P_W_exp}%(32)
\end{equation}
The first and third terms of Eq.~(\ref{eq:P_K_exp}) represent the usual turbulence generation due to the mean electric current and mean velocity strain, respectively.

	As we see in the definitions of $\varepsilon_K$ [Eq.~(\ref{eq:eps_K_def})] and $\varepsilon_W$ [Eq.~(\ref{eq:eps_W_def})], the dissipation rates express how much the turbulent MHD energy and cross helicity are dissipated by the molecular viscosity and magnetic diffusivity in the smallest scales of turbulence. However, in a cascade picture of turbulence, the dissipation rates are tightly connected to the energy and cross-helicity flux in the wave-number space; how much energy and cross helicity are transferred from larger to smaller scales.

	As written in the divergence form, the transport rates $T_K$ and $T_W$ do not contribute to the net generation of $K$ and $W$. However, in some situation, they may play important roles in transport. Since $W$ is non-positive definite, even if the total amount of the cross helicity is zero, it can be spatially distributed as positive and negative pairs. Considering this point, the the first term in Eq.~(\ref{eq:T_W_def}) are expressed in the form of ${\bf{B}} \cdot \nabla K$. This suggests that the turbulent cross helicity is locally generated if the turbulent energy is inhomogeneous along the mean magnetic field. This is a very interesting feature of the cross-helicity generation, which is related to the asymmetry of the Alfv\'{e}n wave propagation.\cite{yok2013}

%-----------------------------------------------------------------------------
%	IV.C Model structure
%-----------------------------------------------------------------------------
\subsection{Model structure}
In this work, we adopt a Reynolds-averaged turbulence model, which consists of the mean-field equations and the turbulence-quantity equations.

%-----------------------------------------------------------------------------
%	IV.C.1 Mean fields
%-----------------------------------------------------------------------------
\subsubsection{Mean fields}
\noindent{\it Mean density:}
\begin{equation}
	\frac{\partial \overline{\rho}}{\partial t}
	+ \nabla \cdot \left( {\overline{\rho} {\bf{U}}} \right)
	= 0,
	\label{eq:mean_den_eq_model}%(33)
\end{equation}

\noindent{\it Mean momentum:}
\begin{eqnarray}
	&& \frac{\partial}{\partial t} \overline{\rho} U^\alpha
	+ \frac{\partial}{\partial x^a} \overline{\rho} U^a U^\alpha
	= - \frac{\partial}{\partial x^\alpha} \overline{\rho} P
	+ \frac{\partial}{\partial x^\alpha} \mu {\cal{S}}^{a \alpha}
	\nonumber\\
	&&\hspace{50pt} 
	+ \overline{\rho} \left ( {{\bf{J}} \times {\bf{B}}} \right)^\alpha
	- \frac{\partial}{\partial x^\alpha} \left( {
		\overline{\rho} {\cal{R}}^{a\alpha}
	} \right),
	\label{eq:mean_mt_eq_model}%(34)
\end{eqnarray}
\noindent{\it Mean energy:}
\begin{eqnarray}
	&&\frac{\partial}{\partial t} \left\{ { \overline{\rho} \left[ {
	\frac{ P}{\gamma_{\rm{s}} - 1}
	+ \frac{1}{2} \left( {{\bf{U}}^2 + {\bf{B}}^2} \right)
	+ \frac{1}{2} \left\langle {
		{\bf{u}}'{}^2  + {\bf{b}}'{}^2
		}\right\rangle
	} \right] } \right\}
	\nonumber\\
	&&\hspace{10pt}= - \nabla \cdot \left\{ {
	\left[ {
	\frac{\gamma_{\rm{s}}}{\gamma_{\rm{s}} - 1} P
	+ \frac{1}{2} \left( {
		{\bf{U}}^2 + \langle {{\bf{u}}'{}^2} \rangle
	} \right)} \right] \overline{\rho} {\bf{U}}
	} \right. \nonumber\\
	&& \left. {
	\hspace{50pt} + \overline{\rho} \left\langle {
	\left( {
	{\bf{U}}\cdot {\bf{u}}'
	} \right) {\bf{u}}'
	} \right\rangle
	+ \overline{\rho} {\bf{E}} \times {\bf{B}} \rule{0.ex}{3.ex}
	} \right\},
	\label{eq:mean_en_eq_model}%(35)
\end{eqnarray}
\noindent{\it Mean magnetic field:}

Faraday induction equation [Eq.~(\ref{eq:mean_faraday_eq})] 

with the mean Ohm's law [Eq.~(\ref{eq:mean_ohms_law})].
\vspace{10pt}

	Here, the Reynolds stress $\mbox{\boldmath${\cal{R}}$}$ and the turbulent electromotive force ${\bf{E}}_{\rm{M}}$ are given by Eqs.~(\ref{eq:rey_stress_exp}) and (\ref{eq:emf_exp}), respectively, with $\alpha$- and $\mbox{\boldmath$\Gamma$}$-related terms dropped. Transport coefficients appearing in Eqs.~(\ref{eq:emf_exp}) and (\ref{eq:rey_stress_exp}) are given by Eq.~(\ref{eq:trans_coeff_rel}).

%-----------------------------------------------------------------------------
%	IV.C.2 Turbulent fields
%-----------------------------------------------------------------------------
\subsubsection{Turbulent fields}
	On the other hand, the equations for the turbulent quantities are as follows.

\noindent{\it Turbulent MHD energy:}
\begin{eqnarray}
	&&\left( {
	\frac{\partial}{\partial t}
	+ {\bf{U}} \cdot \nabla 
	} \right) K
	= - {\bf{E}}_{\rm{M}} \cdot {\bf{J}}
	- {\cal{R}}^{ab} \frac{\partial U^a}{\partial x^b}
	\nonumber\\
	&&\hspace{20pt} - \varepsilon_K
	+ {\bf{B}} \cdot \nabla W
	+ \nabla \cdot \left( {
		\frac{\nu_{\rm{K}}}{\sigma_K} \nabla K
	} \right),
	\label{eq:K_eq_model}%(36)
\end{eqnarray}

\noindent{\it Turbulent cross helicity:}
\begin{eqnarray}
	&&\left( {
	\frac{\partial}{\partial t} + {\bf{U}} \cdot \nabla
	} \right) W
	= - {\bf{E}}_{\rm{M}} \cdot \mbox{\boldmath$\Omega$}
	- {\cal{R}}^{ab} \frac{\partial B^a}{\partial x^b}
	\nonumber\\
	&& \hspace{20pt} - \varepsilon_W
	+ {\bf{B}} \cdot \nabla K
	+ \nabla \cdot \left( {
		\frac{\nu_{\rm{K}}}{\sigma_W} \nabla W
	} \right).	
	\label{eq:W_eq_model}%(37)
\end{eqnarray}

	In order to close the system of model equations, we have to express $\varepsilon$ and $\varepsilon_W$ in terms of known quantities. In the usual Reynolds-averaged turbulence model, in addition to the $K$ equation, the evolution equation of $\varepsilon$ is considered as
\begin{equation}
	\left( {
		\frac{\partial}{\partial t} + {\bf{U}} \cdot \nabla
	} \right) \varepsilon
	= C_{\varepsilon 1}\frac{\varepsilon}{K} P_K
	- C_{\varepsilon 2}\frac{\varepsilon}{K} \varepsilon
	+ \nabla \cdot \left( {
		\frac{\nu_{\rm{K}}}{\sigma_\varepsilon} \nabla \varepsilon
	} \right),
	\label{eq:eps_eq_model}%(38)
\end{equation}
where $C_{\varepsilon 1}$, $C_{\varepsilon 2}$, and $\sigma_{\varepsilon}$ are the model constants, whose values have been optimized through the several kinds of flow environments as\cite{lau1972}
\begin{equation}
	C_{\varepsilon 1} = 1.4,\;
	C_{\varepsilon 2} = 1.9,\;
	\sigma_\varepsilon = 1.3. 
	\label{eq:eps_eq_model_const}%(39)
\end{equation}
These values should be kept the same even in the MHD turbulence case since a system of model equations should be reduced to the usual hydrodynamic turbulence model in the limit of vanishing magnetic field (${\bf{b}}=0$). The equation for $\varepsilon_W$ can be also constructed.\cite{yok2011a} A simpler modeling of $\varepsilon$ and $\varepsilon_W$ is the algebraic model for them. By introducing timescale of turbulence, $\tau$, $\varepsilon$ and $\varepsilon_W$ are modeled as
\begin{equation}
	\varepsilon = \frac{K}{\tau},
	\label{eq:eps_K_alg_model}%(40)
\end{equation}
\begin{equation}
	\varepsilon_W = C_W \frac{W}{\tau}
	\label{eq:eps_W_alg_model}%(41)
\end{equation}
[$C_W (>1)$ is the model constant]. We should note that in the latter case the system of model equations is not entirely closed since the timescale $\tau$ can not be self-consistently determined without resorting to the $\varepsilon$ equation. As we see from Eq.~(\ref{eq:eps_K_alg_model}), only when we solve both $K$ and $\varepsilon$ equations simultaneously, the timescale of turbulence is self-consistently determined as
\begin{equation}
	\tau = \frac{K}{\varepsilon}.
	\label{eq:timescale_model}%(42)
\end{equation}
The detailed modeling of the $\varepsilon$ and $\varepsilon_W$ including the derivation of their equations, has been given in the previous papers.~\cite{yos1987,yok2011a}

%-----------------------------------------------------------------------------
%	V Notes on mean-field approach
%-----------------------------------------------------------------------------
\section{Notes on mean-field approach}
We adopt the mean-field approach to the magnetic reconnection in the present work. Validity of this approach should be examined in many respects. In this section, we discuss several points related to the mean-field approach, which include the applicability to the magnetic reconnection phenomena, assumptions behind the approach, advantages and disadvantages, potentiality for the large-eddy simulations (LESs), and test of the mean-field approach.

%-----------------------------------------------------------------------------
%	V.A Applicability of the mean-field approach
%-----------------------------------------------------------------------------
\subsection{Applicability of the mean-field approach}
Richardson considered the dispersion of particle pairs passively advected by homogeneous, isotropic and fully developed turbulence.\cite{ric1926} Ensemble averaging of such advected fields gives the notion of stochastic diffusion due to turbulence (Richardson diffusion). Mean-field approaches to turbulence with effective transports should be based explicitly or implicitly on this notion. As Eyink {\it et al.} pointed out, the Lagrange trajectory problem is equivalent to solving the initial value problem of the magnetic induction equation.\cite{eyi2011} Ensemble of the magnetic-field lines subject to the Richardson diffusion exhibits a fast topology change, which corresponds to the fast reconnection. This indicates that turbulence should involve the fast reconnection on the scale of its eddies. 

	Bearing this point in mind, we may say that the mean-field approach should be based on or at least compatible with the fast reconnection of the small-scale magnetic field. This is a very interesting aspect of the mean-field approach to magnetic reconnection. 
	The compatibility of the mean-field approach with the small-scale reconnection can be confirmed only through the examination of the turbulence model [Eqs.~(\ref{eq:emf_exp}) and (\ref{eq:rey_stress_exp})] with the aid of the direct numerical simulations (DNSs) of the small-scale reconnection phenomena. Although the validity of the mean-field turbulence model has been confirmed in some magnetohydrodynamic flow situations,\cite{yok2011c} the general confirmations of the model is lacking in the context of small-scale magnetic reconnection. Again we should note that the numerical simulations with a mean-field turbulence model do not provide us with a direct test of the turbulence reconnection model.

	In the mean-field approach, only the mean (ensemble averaged) motions are treated as the resolved or calculated ones, whereas all the turbulent motions (deviations from the mean) are treated as the unresolved ones, which should be modeled. Here, it is important to note that conceptionally the ensemble averaged fields are not necessarily the large-scale fields. The equations for the turbulent motions are replaced by a reduced set of equations. In order to construct the reduction theory, we adopt a few quantities that represent the statistical properties of turbulent motions. The transport coefficients appearing in the models for the turbulent correlations such as the turbulent EMF, the Reynolds stress, etc.\ are expressed in terms of the turbulent statistical quantities and timescales of turbulence as in Eq.~(\ref{eq:trans_coeff_rel}). As for the representative turbulent statistical quantities, we often adopt one-point turbulent quantities such as the turbulent (MHD) energy, its dissipation rate, the turbulent kinetic helicity, the turbulent magnetic helicity, the turbulent cross helicity, etc. These one-point statistical quantities, which are equivalent to the spectral integrals of the corresponding turbulent spectral functions, do not contain any information on scale dependence. Since most energy are attributed to the largest-scale motions, it is obvious that the magnitudes of the turbulent statistical quantities are mainly determined by the largest-scale motions of turbulence. In this sense, turbulent transport coefficients appearing in the mean-field equations are introduced on the basis of the largest or energy-containing motions in turbulence. Practically this is often the case with the mean-field approach, and this point matches the finding by Eyink {\textit{et al.}}\ (2011) that a formal introduction of magnetic diffusivity is possible only for the systems much larger than the energy injection scale.\cite{eyi2011} In order to treat the motions much smaller than the injection scale, the mean-field approach employing one-point quantities may be not sufficient, and we have to treat the spectral information of turbulence in a more elaborated manner.

%-----------------------------------------------------------------------------
%	V.B Suitability of the mean-field approach to treat realistic turbulence
%-----------------------------------------------------------------------------
\subsection{Suitability of the mean-field approach to treat realistic turbulence}
	Provided that the turbulence model is appropriate enough,  mean-field approaches are suitable for treating the large-scale magnetic reconnection with realistic parameters. This is because in the mean-field approach we can easily treat turbulence at huge kinetic and magnetic Reynolds numbers including the huge Lundquist number. In the approach, all the turbulent motions are reduced to a small number of turbulence statistical quantities. The validity of the mean-field approach highly depends on the turbulence model we adopt. The model should properly represent the unresolved motions or reduced information of turbulence. In particular in the case of turbulent reconnection, the expressions for the turbulent magnetic diffusivity is of primary importance. As is well known, simplest representations such as the mixing-length model is not good enough. 
	
	In the current theory of turbulent reconnection,\cite{eyi2011} the eddy diffusivity is estimated with the correlation method for the Lagrangian fluid particle as
\begin{equation}
	D_{\rm{T}}^{\alpha\beta}
	= \frac{1}{2} \int_{-\infty}^{\infty} \!\!d\tau \left\langle {
		V^\alpha(\tau) V^\beta(0)
	} \right\rangle
	\label{eq:D_T_exp}%(43)
\end{equation}
(${\bf{V}}(t) = d{\bf{x}}(t)/dt$: Lagrangian fluid particle velocity). On the basis of Eq.~(\ref{eq:D_T_exp}), the turbulent or eddy diffusivity for the Alfv\'{e}n wave turbulence is estimated using scaling arguments. There the nonlinear Alfv\'{e}n wave turbulence timescale is expressed as $\tau \sim M_{\rm{A}}^{-2} \omega_{\rm{A}}^{-1}$ [$M_{\rm{A}} (= u_L / V_{\rm{A}})$: Alfv\'{e}n Mach number, $\omega_{\rm{A}} (= V_{\rm{A}}/L_{\rm{i}})$: forcing-scale Alfv\'{e}n wave frequency]. 

	On the other hand, in the present framework, the turbulent magnetic diffusivity $\beta$ is estimated based on
\begin{eqnarray}
	\beta &=& \int \!\!d{\bf{k}} \int_{-\infty}^t \!\!d\tau_1
		G(k, {\bf{x}}; \tau, \tau_1) \times
	\nonumber\\
		& & \hspace{10pt} \left[ {
			Q_{uu}(k, {\bf{x}}; \tau, \tau_1, t) 
			+ Q_{bb}(k, {\bf{x}}; \tau, \tau_1, t)
		} \right],
	\label{eq:beta_spect_exp}%(44)
\end{eqnarray}
where $Q_{uu}$, $Q_{uu}$, and $G$ are the kinetic and magnetic energy spectral functions and the response function of turbulence. [For the assumptions and approximations in deriving Eq.~(\ref{eq:beta_spect_exp}), the reader is referred to Appendix of Ref.~\onlinecite{yok2013} and references cited therein.] Equation~(\ref{eq:beta_spect_exp}) indicates that the turbulent magnetic diffusivity is determined by the intensity of fluctuations and the timescale related to how much the past state affects the present one. If the energy spectra can be treated as independent of the past history, Eq.~(\ref{eq:beta_spect_exp}) just gives $\beta \sim K \tau \sim u \ell$ [$K (\sim u^2)$: energy, $\ell$: mixing length]. This is a simple mixing-length expression for $\beta$. Equation~(\ref{eq:beta_spect_exp}) can be regarded as an elaborated generalization of the mixing-length model.

	Unlike the simple mixing-length model, where the mixing length is prescribed by some typical length scale such as the density or pressure scale hight, in the present work, the turbulent magnetic diffusivity [Eq.~(\ref{eq:beta_nuk_rel})] is determined by solving the transport equations of the turbulence statistical quantities. The equation of turbulent MHD energy $K$ [Eq.~(\ref{eq:K-W_eq})] is solved with Eq.~(\ref{eq:K_eq_terms}), which contains the turbulence generation mechanism fully related to the mean-field inhomogeneity such as the mean electric-current density ${\bf{J}}$ and the mean velocity shear $\partial U^a / \partial x^b$. Note that if turbulence is homogeneous, we have no turbulence production mechanisms represented by Eqs.~(\ref{eq:K_eq_terms}) and (\ref{eq:W_eq_terms}) at all. In such a situation, we need external source of turbulence represented by external forcing to generate and sustain turbulence. It is this situation that most previous turbulence reconnection studies had considered.\cite{mat1985,mat1986,laz1999,eyi2011} We should remember that, in the framework of Goldreich \& Sridhar theory,\cite{gol1995} on which several turbulent reconnection studies are based,\cite{laz1999,eyi2011} it is assumed that the mean magnetic field is uniform (${\bf{B}} = {\bf{B}}_0$) and the mean velocity is completely absent (${\bf{U}} = 0$).

	Note also that Eq.~(\ref{eq:beta_spect_exp}) is derived by an analysis of a full set of magnetohydrodynamic (MHD) equations (not only the magnetic induction equation). In order to fully appreciate Eq.~(\ref{eq:beta_spect_exp}), we need information of the Green's function. Actually, the Green's function tensor $G^{\alpha\beta}$, from which $G$ in Eq.~(\ref{eq:beta_spect_exp}) arises, obeys the wave-number space fluctuation equation: 
\begin{eqnarray}
	& & \frac{\partial G_\phi^{\alpha\beta}({\bf{k}}; \tau, \tau')}{\partial \tau}
	+ \nu k^2 G_\phi^{\alpha\beta}({\bf{k}}; \tau, \tau')
	\nonumber\\
	& & \hspace{10pt} - i Z^{\alpha ab}({\bf{k}}) 
	\int \!\!d{\bf{p}}d{\bf{q}}\ \delta({\bf{k}} - {\bf{p}} - {\bf{q}}) 
	\psi'{}^a({\bf{p}}; \tau) G_\phi^{b\beta}({\bf{q}}; \tau, \tau')
	\nonumber\\
	& & \hspace{10pt} = \delta^{\alpha\beta} \delta(\tau - \tau')
	\label{eq:Green_fn_eq}%(45)
\end{eqnarray}
where $\mbox{\boldmath$\phi$} (= {\bf{u}} + {\bf{b}})$ and $\mbox{\boldmath$\psi$} (= {\bf{u}} - {\bf{b}})$ are Elsasser variables and $Z^{\alpha ab}({\bf{k}}) = k^a D^{\alpha b}({\bf{k}})$ with the projection operator $D^{\alpha\beta}({\bf{k}}) = \delta^{\alpha\beta} - k^\alpha k^\beta / k^2$. Equation~(\ref{eq:Green_fn_eq}) is the MHD counterpart of the response-function equation in hydrodynamic turbulence.\cite{kra1959} At a high Reynolds number, the third term in the left-hand side of Eq.~(\ref{eq:Green_fn_eq}) plays a dominant role. As the spectral integral of the third term shows, the dynamics of ${\bf{k}}$ mode is determined by the spectral interaction with all the other modes ${\bf{p}}$ and ${\bf{q}}$. This suggests that, in order to test the present mean-field approach in a fundamental manner, we have to evaluate the Green's functions of inhomogeneous turbulence with the aid of the direct numerical simulations (DNSs). This is a very interesting but challenging problem.

	The validation of the mean-field approach can be performed at less fundamental levels. In the present work, we choose a combination of statistical quantities, such as the turbulent energy $K$, its dissipation rate $\varepsilon$, cross helicity $W$, etc., on the basis of the expressions obtained from the closure theory of inhomogeneous MHD turbulence. For example, the turbulent magnetic diffusivity is expressed as Eq.~(\ref{eq:beta_spect_exp}), which is modeled with the turbulent MHD energy $K$ as Eq.~(\ref{eq:beta_nuk_rel}). These statistical quantities are expected to properly represent the statistical properties of turbulence. Then we solve the transport equations for these statistical quantities in addition to the mean-field equations. Effects of mean-field inhomogeneities are taken into account through the production and transport rates in the transport equations. 

	It is obvious that the expressions [Eq.~(\ref{eq:trans_coeff_rel})] for the transport coefficients in terms of the one-point turbulence quantities should be examined with the aid of the DNSs. Estimate of the timescales in comparison with the Green's function is important. The expressions of the Reynolds stress [Eq.~(\ref{eq:rey_stress_exp})] and the turbulent electromotive force [Eq.~(\ref{eq:emf_exp})] should be also checked.

%-----------------------------------------------------------------------------
%	V.C Potentiality for large-eddy simulations (LESs)
%-----------------------------------------------------------------------------
\subsection{Potentiality for large-eddy simulations (LESs)}
One feature of large-eddy simulations (LESs) lies in the point that it is suitable for treating unsteady phenomena. In order to examine the unsteady phases of magnetic reconnection represented by several instabilities, such a feature of LESs must be very favorable. On the other hand, the mean-field approach with the Reynolds- or ensemble-averaged turbulence model is suitable for treating a steady state without resorting a high cost of numerical calculations. Models for the unresolved- or subgrid-scale (SGS) motions are often constructed under the assumption that there is a local equilibrium between the SGS energy and its transfer rate from larger to smaller scales. In any case, we have to validate the SGS model with the aid of the direct numerical simulations (DNSs). Such validations may include the estimate of how much SGS energy is generated from the small-scale reconnection itself, how much SGS energy is cascaded from larger to smaller scales through the grid-scale (GS) inhomogeneities, etc.

%-----------------------------------------------------------------------------
%	V.D Correspondence to homogeneous turbulence
%-----------------------------------------------------------------------------
\subsection{Correspondence to homogeneous turbulence}
	It is important to consider how our mean-field approach is related to the previous reconnection models based on homogeneous turbulence. In one sense, it may be difficult to make such correspondence since turbulence generation mechanisms are fairly different between these two approaches. In homogeneous turbulence approach, we need some energy sources that are externally driven by forcing or that is intrinsic to the small-scale turbulence itself. In the inhomogeneous turbulence approach the source of turbulence energy are provided by cascades from the mean-field energy through the large-scale inhomogeneities. In order to investigate how these two approaches can be related to each other, we need direct numerical simulations (DNSs) in a situation with mean-field inhomogeneities and check how much energy is transferred between the mean fields and turbulence. 

	This immediately suggests that the key is how to formulate the ``mean'' fields in a practical calculation. Theoretically, mean can be just formulated as the ensemble average, but in practical simulations, such formulation is not so simple. One possible candidate is provided by a previous work.\cite{yok2011c} In the work, we considered the Kolmogorov flow with a uniform magnetic field imposed, and examined the turbulent electromotive force (EMF) $\langle {{\bf{u}}' \times {\bf{b}}'} \rangle$ and compared its spatial distribution with the mean-field model expressions for the EMF. The Kolmogorov flow has one inhomogeneous direction whereas the statistical quantities are homogenous in the other two directions (horizontal directions). We can formulate the mean fields by using the horizontal (and some appropriate time) averaging. In the context of the magnetic reconnection, a three-dimensional direct numerical simulation (DNS) with one homogeneous direction may provide such a situation. In this case, we can formulate the mean fields by the fields averaged in the homogeneous direction.

%-----------------------------------------------------------------------------
%	VI Numerical test
%-----------------------------------------------------------------------------
\section{Numerical test}
Recently a system of mean-field equations with turbulence effect incorporated has been investigated in the context of the magnetic reconnection.\cite{yok2011b} In the work, stationary solutions for the mean-field equations are analytically derived with the turbulence effects. It is predicted that the reconnection is confined to a tiny region due to the dynamic balance between the enhanced magnetic diffusivity due to the turbulence energy and the transport suppression due to the turbulent cross helicity. These notions should be numerically or experimentally confirmed. For this purpose, we perform a numerical simulation of magnetic reconnection with turbulence effect self-consistently implemented through a turbulence model.

	Here we should remark the following points. This is a numerical simulation based on a turbulence model. The results of the simulation inevitably depend on the turbulence model we adopt. In this sense, the simulation results can not provide any direct test for the reconnection model. In order to make a direct test, we need direct numerical simulations (DNSs) of the turbulent magnetic reconnection. There have been several elaborated DNSs that contribute to the understanding of the turbulent reconnection.\cite{kow2009,lou2009,kow2012} However, such DNSs with astrophysical realistic parameters are impossible in the foreseeable future. On the contrary, the mean-field or turbulence model approach provides a powerful tool for investigating the turbulent magnetic reconnection with the effects of strongly sheared large-scale fields and fast flow along the magnetic fields in the realistic physical parameters.

%-----------------------------------------------------------------------------
%	VI.A Equations to solve
%-----------------------------------------------------------------------------
\subsection{Equations to solve}
	The equations of the mean fields and turbulence numerically to solve in this work are as follows. 
\begin{equation}
	\frac{\partial \overline{\rho}}{\partial t}
	= - \nabla \cdot \left( {\overline{\rho} {\bf{U}}} \right),
	\label{eq:den_eq_num}%(46)
\end{equation}
\begin{equation}
	\frac{\partial}{\partial t} \left( {\overline{\rho} {\bf{U}}} \right)
	= - \nabla \cdot \left\{ { \overline{\rho} \left[ {
		\left( {
			{\bf{U}} {\bf{U}} - {\bf{B}} {\bf{B}}
		} \right)
		+ \left( {P + \frac{1}{2} {\bf{B}}^2}\right) {\mbox{\boldmath$\cal{I}$}}
	} \right] } \right\},
	\label{eq:vel_eq_num}%(47)
\end{equation}
\begin{eqnarray}
	&&\frac{\partial}{\partial t} \left\{ {
		 \overline{\rho} \left[ {
			\frac{P}{\gamma_{\rm{s}} - 1}
			+ \frac{1}{2} \left( {{\bf{U}}^2 + {\bf{B}}^2} \right)
		} \right]
	} \right\}
	\nonumber\\
	&&\hspace{0pt}= - \nabla \cdot \left\{ {  \overline{\rho}
		\left[ {
		\left( {
			\frac{\gamma_{\rm{s}}}{\gamma_{\rm{s}} - 1} P
			+ \frac{1}{2} {\bf{U}}^2
		} \right) {\bf{U}}
	+ {\bf{E}} \times {\bf{B}}
	} \right]
	} \right\},
	\label{eq:energy_eq_num}%(48)
\end{eqnarray}
\begin{equation}
	\frac{\partial {\bf{B}}}{\partial t}
	= - \nabla \times {\bf{E}},
	\label{eq:mean_faraday_num}%(49)
\end{equation}
\begin{equation}
	{\bf{E}}
	= - {\bf{U}} \times {\bf{B}}
	+ \eta {\bf{J}}
	- \tau \left( {
		C_\gamma W \mbox{\boldmath$\Omega$} - C_\beta K {\bf{J}}
	} \right),
	\label{mean_ohm_num}%(50)
\end{equation}
\begin{equation}
	\frac{\partial K}{\partial t}
	= - {\bf{U}} \cdot \nabla K
	+C_\beta  \tau K {\bf{J}}^2
	- C_\gamma \tau W \mbox{\boldmath$\Omega$} \cdot {\bf{J}}
	+ {\bf{B}} \cdot \nabla W
	- \frac{K}{\tau},
	\label{eq:K_eq_num}%(51)
\end{equation}
\begin{equation}
	\frac{\partial W}{\partial t}
	= - {\bf{U}} \cdot \nabla W
	+ C_\beta \tau K \mbox{\boldmath$\Omega$} \cdot {\bf{J}}
	- C_\gamma \tau W \mbox{\boldmath$\Omega$}^2
	+ {\bf{B}} \cdot \nabla K
	- C_W \frac{W}{\tau},
	\label{eq:W_eq_num}%(52)
\end{equation}
where $\mbox{\boldmath$\cal{I}$}$ is the unit tensor, and the dynamic viscosity $\mu$ is neglected. In Eqs.~(\ref{eq:vel_eq_num}), (\ref{eq:energy_eq_num}), (\ref{eq:K_eq_num}) and (\ref{eq:W_eq_num}), the Reynolds stress-related terms and transport rate terms other than ${\bf{B}} \cdot \nabla W$ and ${\bf{B}} \cdot \nabla K$ are dropped. As for the dissipation rates of $K$ and $W$, $\varepsilon$ and $\varepsilon_W$, we adopt algebraic expressions for them [Eqs.~(\ref{eq:eps_K_alg_model}) and (\ref{eq:eps_W_alg_model})] instead of using the transport equations for $\varepsilon$ [Eq.~(\ref{eq:eps_eq_model})] and $\varepsilon_W$.\cite{yok2011a} As for the model constants, we adopt
\begin{equation}
	C_\beta = C_\gamma = 0.3,\;\;
	C_W = 1.3
	\label{eq:model_consts}%(53)
\end{equation}
in this work.\cite{yok2008} The change of these values does not lead to a qualitative difference in the numerical results.

%-----------------------------------------------------------------------------
%	VI.B Set-up
%-----------------------------------------------------------------------------
\subsection{Set-up}
The above set of equations is solved by the two-dimensional ($x$-$y$ plane) 4th-order Runge--Kutta and 4th order central difference scheme. The grid intervals $\Delta x$ and $\Delta y$ are fixed to unity and the simulation size is $L_x \times L_y = 2048 \times 512$. The periodic boundary conditions are assumed in both $x$ and $y$ directions for simplicity.

	As for the initial fields, we assume a pair of Harris sheets in the $x$-$y$ plane. The profile of the mean magnetic field is given as
\begin{eqnarray}
	{\bf{B}} &=& {\bf{e}}_x B_{x0} \left[ {
	\tanh \left( {\frac{y}{\delta}} \right)
	- \tanh \left( {\frac{y - 0.5 L_y}{\delta}} \right)
	-1
	} \right]
	\nonumber\\
	&+&{\bf{e}}_y B_{y0} \sum_{m=1}^{10} 
	\sin \left( {
		\frac{2\pi mx}{L_x}
	} \right)
	\label{eq:initial_B}%(54)
\end{eqnarray}
with the small magnetic field perturbation in order to initiate reconnection. Here $B_{y0}/B_{x0} = 1.0 \times 10^{-3}$ and $\delta = 0.02 L_y$ (${\bf{e}}_x$ and ${\bf{e}}_y$ are the unit vectors in the $x$ and $y$ directions, respectively). The plasma beta outside the current sheet is set to be $\beta_{\rm{p}} = 0.5$. The initial turbulent MHD energy is put $K = 5.0 \times 10^{-3}$ homogeneously in space. The initial mean velocity and the turbulent cross helicity are set equal to zero (${\bf{U}} = 0, W=0$). For details of set-up for numerical simulation, see the other report.\cite{hig2013a}

%-----------------------------------------------------------------------------
%	VI.C Results
%-----------------------------------------------------------------------------
\subsection{Results}
\begin{figure}[htb]
\includegraphics[width=.45\textwidth]{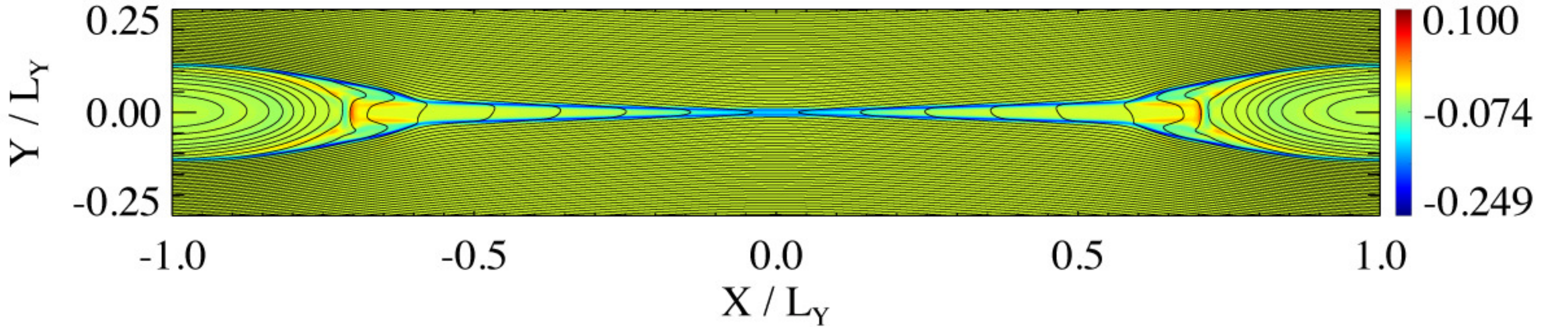}
\caption{\label{fig:el_crrnt} Spatial distribution of the mean electric-current density $J^z$.}
\end{figure}

The detailed results of numerical simulation are reported in another paper, which includes the temporal evolution of the mean-field structures, the reconnected magnetic fluxes, the budget of the turbulence transport equations, and their dependence on the turbulence dissipation rate, etc.\cite{hig2013a} Basic results are in agreement with the theoretical predictions.\cite{yok2011b} The mean-field configuration favorable for the fast reconnection is realized as a steady state (Fig.~\ref{fig:el_crrnt}). Turbulence is self-generated and sustained by the mean-field configurations, the reconnection rate is drastically enhanced by turbulence. 

	As has been referred to in \S~1, one interesting point is that too much high level of fluctuation results in not fast reconnections but turbulent diffusions of the mean magnetic fields. In this case, the mean-field configurations favorable for the magnetic reconnection, the oppositely directed magnetic fields, themselves disappear due to the high turbulent magnetic diffusivity, and the reconnection rate based on the reconnected magnetic fields is even lower than the slow or laminar reconnection case (see figures 1-3 of Ref.~\onlinecite{hig2013a}).

	Related to this, we should note the following point. In the present numerical simulations, the dissipation rate of turbulent energy, $\varepsilon$, is estimated by using an algebraic model $\varepsilon = K / \tau = K / (C_\tau \tau_0)$ with $\tau_0$ and $C_\tau$ being the initial timescale of turbulence and model constants, respectively. In other words, the level of turbulence is controlled by the energy dissipation rate through the time-scale constant $C_\tau$ in the present simulations. This is an artifact since the dissipation rate of the  turbulent energy should be determined by the nonlinear dynamics of turbulence itself. In order to fully treat this dynamics in the turbulence model simulations, we have to solve the energy dissipation-rate equation as well as the turbulence energy equation.\cite{yok2008,yok2011a} With this reservation in mind, our numerical results are still suggestive in showing the role of turbulent magnetic diffusivity in the context of the mean or large-scale magnetic-field reconnection.

	Hereafter we confine ourselves to the results concerning the dynamic balance between the enhanced magnetic diffusivity due to turbulence and the cross-helicity-related transport suppression.

%------------------------------------------
%	Fig. 2
%------------------------------------------

\begin{figure}[htb]
\includegraphics[width=.45\textwidth]{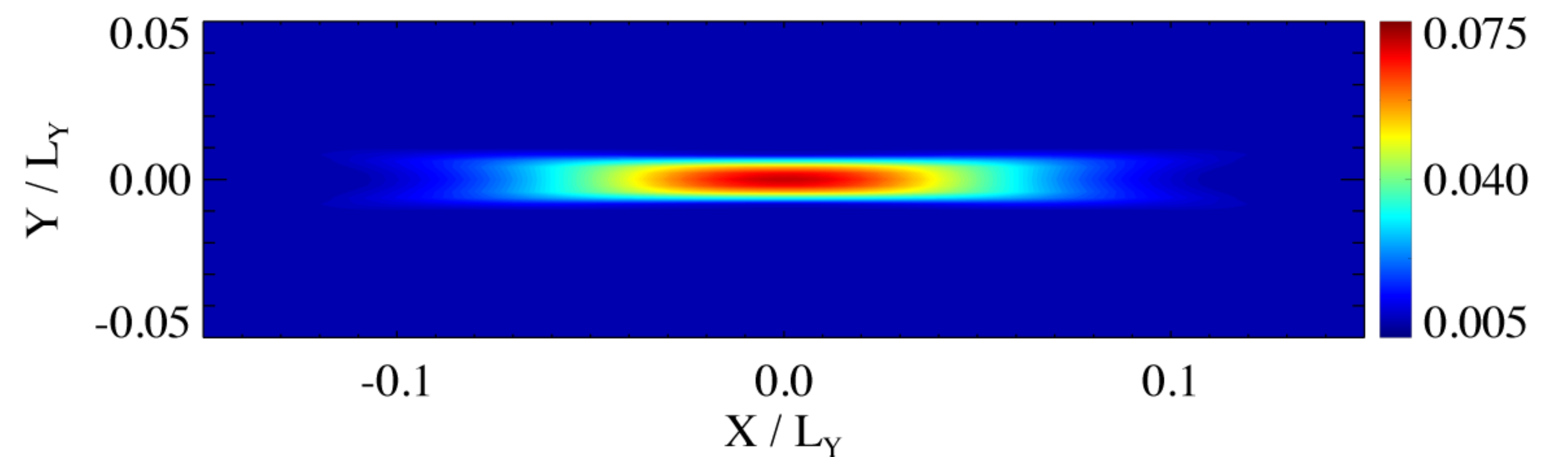}
\caption{\label{fig:turb_en} Spatial distribution of the turbulent MHD energy $K = \langle {{\bf{u}}'{}^2 + {\bf{b}}'{}^2} \rangle/2$ around the magnetic reconnection region.}
\end{figure}

	The spatial distributions of the turbulent MHD energy $K$ and the turbulent cross helicity $W$ are shown in Figs.~\ref{fig:turb_en}-4. Figure~\ref{fig:turb_en} shows that the turbulent MHD energy is spatially distributed in the vicinity of the reconnection point. As we see from the second term in the r.h.s.\ of Eq.~(\ref{eq:K_eq_num}), one of the main production mechanisms of turbulence is $\beta {\bf{J}}^2$. Since the mean electric-current density is strongest there, the turbulent MHD energy $K$ is largest in the vicinity of the reconnection point. The most important point here is that the level of turbulence and its spatiotemporal distribution are dynamically determined by the coupling of the mean-field configuration and turbulence. In this sense, this turbulence is self-generated and sustained by the dynamics of the magnetic reconnection environment.

%------------------------------------------
%	Fig. 3
%------------------------------------------

\begin{figure}[htb]
\includegraphics[width=.45\textwidth]{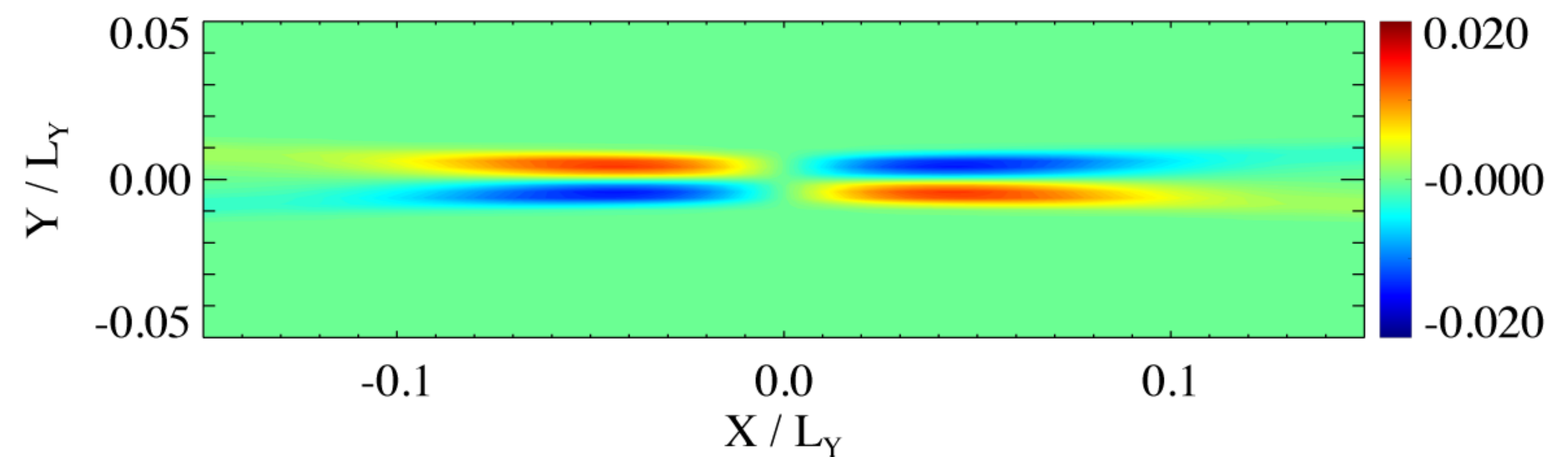}
\caption{\label{fig:turb_ch} Spatial distribution of the turbulent cross helicity $W = \langle {{\bf{u}}' \cdot {\bf{b}}'} \rangle$ around the magnetic reconnection.}
\end{figure}

	Figure~\ref{fig:turb_ch} shows that the cross helicity is spatially distributed in a quadrupole manner with respect to the reconnection point. This result is completely in agreement with the theoretical prediction.\cite{yok2011b} Note that we do not presume any turbulent cross helicity at all in the present numerical simulation. The cross helicity is spontaneously generated with the mean velocity by the inhomogeneities of the mean fields. Its emergence and persistent presence are just a consequence of the symmetric and antisymmetric property of the mean fields associated with the magnetic reconnection. This spatial distribution is mainly caused by the second term in the r.h.s.\ of Eq.~(\ref{eq:W_eq_num}), which suggests that the sign of the cross-helicity generation is determined by the sign of $\mbox{\boldmath$\Omega$} \cdot {\bf{J}}$. The direction of the mean electric-current density ${\bf{J}}$ is the same everywhere whereas that of the mean vorticity $\mbox{\boldmath$\Omega$}$ changes depending on the directions of the mean flow with respect to the mean magnetic field. This quadrupole distribution of the turbulent cross helicity must be fairly ubiquitous around the reconnection point since it arises just from the symmetric and antisymmetric properties of the magnetic and velocity fields.

%------------------------------------------
%	Fig. 4
%------------------------------------------

\begin{figure}[htb]
\includegraphics[width=.45\textwidth]{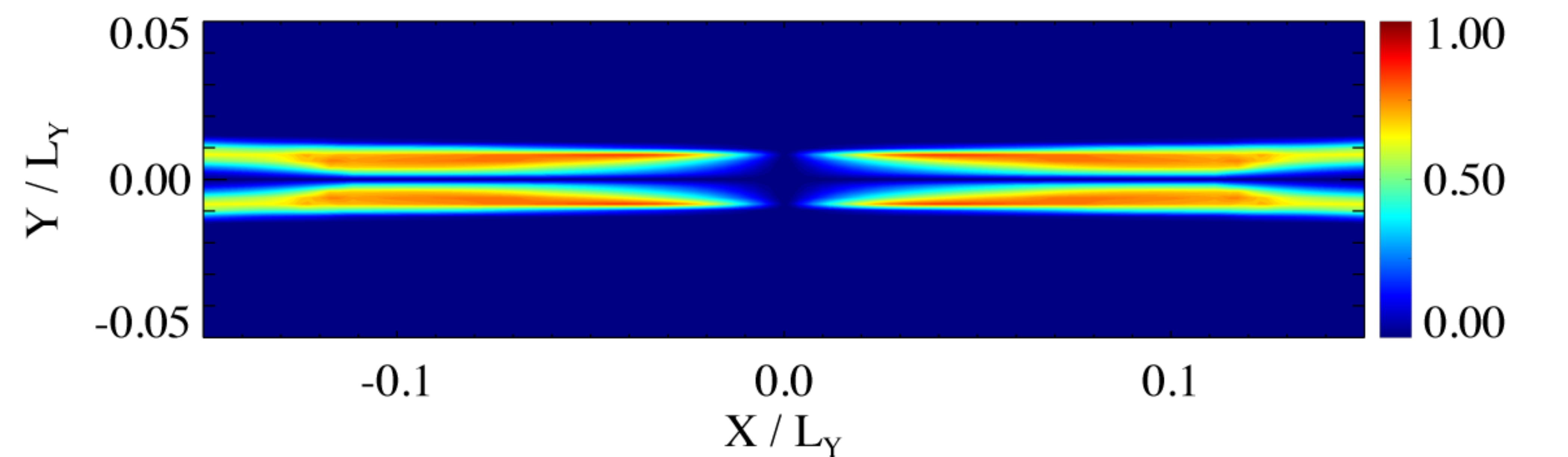}
\caption{\label{fig:turb_ch_over_en} Spatial distribution of the scaled magnitude of the turbulent cross helicity $|W|/K = 2 |\langle {{\bf{u}}' \cdot {\bf{b}}'} \rangle| / \langle {{\bf{u}}'{}^2 + {\bf{b}}'{}^2} \rangle$ around the magnetic reconnection region.}
\end{figure}

	Figure~\ref{fig:turb_ch_over_en} shows the spatial distributions of the turbulent cross helicity scaled by the turbulent MHD energy:
\begin{equation}
	\frac{|W|}{K} 
	= \frac{|\langle {
		{\bf{u}}' \cdot {\bf{b}}'
	} \rangle|}{\langle {
		{\bf{u}}'{}^2 + {\bf{b}}'{}^2
	} \rangle/2}.
	\label{eq:scaled_cross_hel}%(55)
\end{equation}
This is one of the most important non-dimensional quantities in the cross-helicity-related dynamos and transport phenomena. The magnitude of this quantity around the reconnection point is of $O(10^{-1})$, which is remarkably high compared with the other situations. This large value results from the presence of strong magnetic fields in the reconnection situation and the global breakage of symmetry between the directions parallel and antiparallel to the mean magnetic field.

%------------------------------------------
%	Fig. 5
%------------------------------------------

\begin{figure}[htb]
%\begin{tabular}{c}
%\subfloat[]{\includegraphics[width=.35\textwidth]{n_yokoi_mr2012_fig_05a}}\\
%\subfloat[]{\includegraphics[width=.35\textwidth]{n_yokoi_mr2012_fig_05b}}\\
%\subfloat[]{\includegraphics[width=.35\textwidth]{n_yokoi_mr2012_fig_05c}}\\
%\end{tabular}
\includegraphics[width=.45\textwidth]{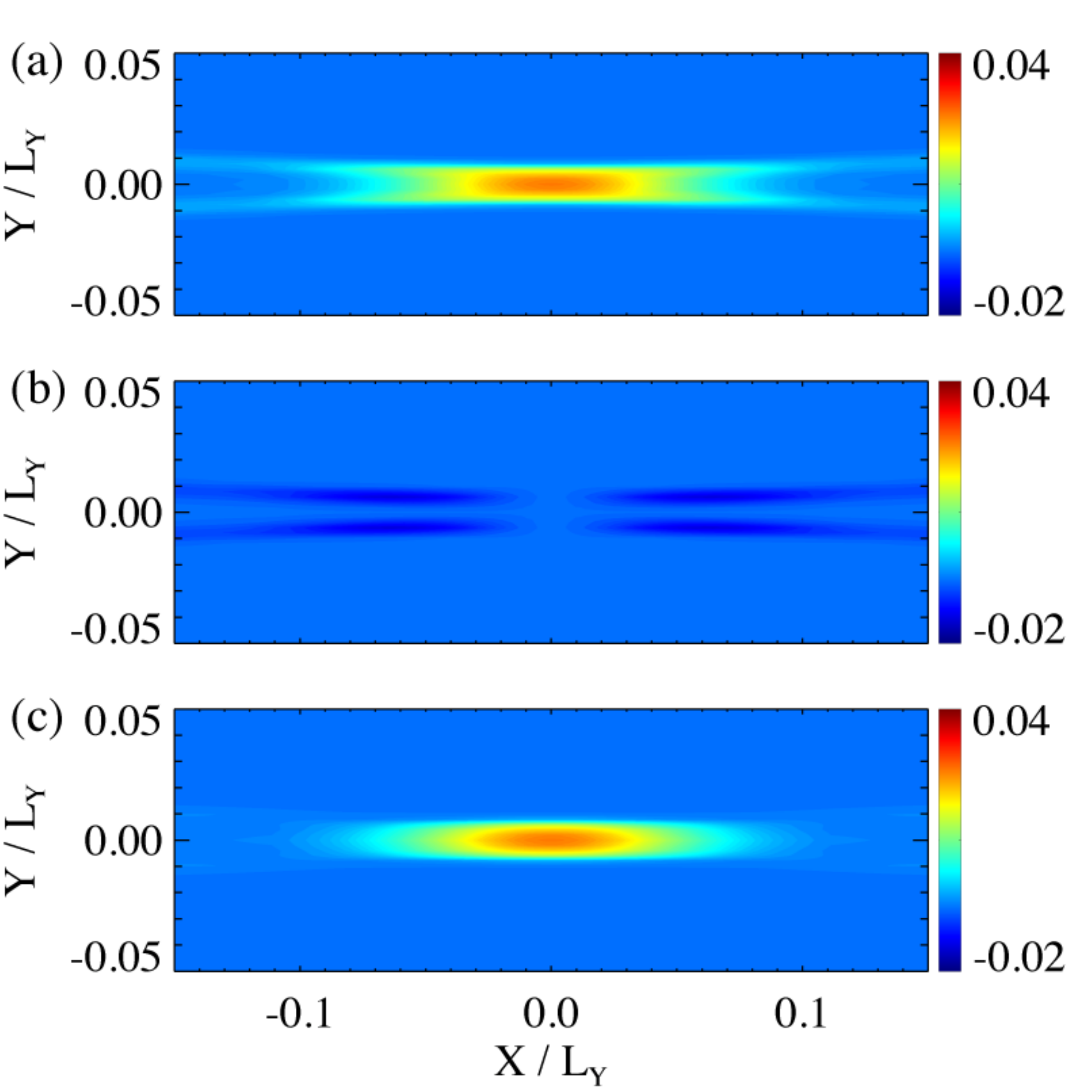}
\caption{\label{fig:turb_confinement} Spatial distributions of (a) $-\beta J^z$, (b) $\gamma \Omega^z$, and (c) $- \beta J^z + \gamma \Omega^z$ around the magnetic reconnection region.}
\end{figure}

	The dynamic balance between the transport enhancement due to turbulent energy and the suppression due to the turbulent cross helicity is directly expressed by the residual of these two effects, which is expressed by the r.h.s.\ of Eq.~(\ref{eq:emf_exp}) with the $\alpha$-related term dropped:
\begin{eqnarray}
	{\bf{E}}_{\rm{M}} 
	&=& - \beta {\bf{J}} + \gamma \mbox{\boldmath$\Omega$}
	\nonumber\\
	&=& - C_\beta \tau K {\bf{J}} + C_\gamma \tau W \mbox{\boldmath$\Omega$}.
	\label{eq:dyn_balance}%(56)
\end{eqnarray}
Figure~\ref{fig:turb_confinement} shows the spatial distributions of the $z$ component of $-\beta {\bf{J}}$, $\gamma \mbox{\boldmath$\Omega$}$, and $- \beta {\bf{J}} + \gamma \mbox{\boldmath$\Omega$}$. If we compare the spatial distribution of $- \beta {\bf{J}}$ with the counterpart of $- \beta {\bf{J}} + \gamma \mbox{\boldmath$\Omega$}$, the contrast in the latter [Fig.~\ref{fig:turb_confinement}(c)] is stronger than the former [Fig.~\ref{fig:turb_confinement}(a)]. This means that, because of the dynamic balance between the transport enhancement due to the turbulent energy $\beta {\bf{J}}$ and the transport suppression due to the turbulent cross helicity $\gamma \mbox{\boldmath$\Omega$}$, the effective magnetic diffusion is localized in a narower region. This contributes to the fast reconnection.

%------------------------------------------
%	Fig. 6
%------------------------------------------

\begin{figure}[htb]
%\begin{tabular}{c}
%\subfloat[]{\includegraphics[width=.35\textwidth]{n_yokoi_mr2012_fig_06a}}\\
%\subfloat[]{\includegraphics[width=.35\textwidth]{n_yokoi_mr2012_fig_06b}}\\
%\end{tabular}
\includegraphics[width=.45\textwidth]{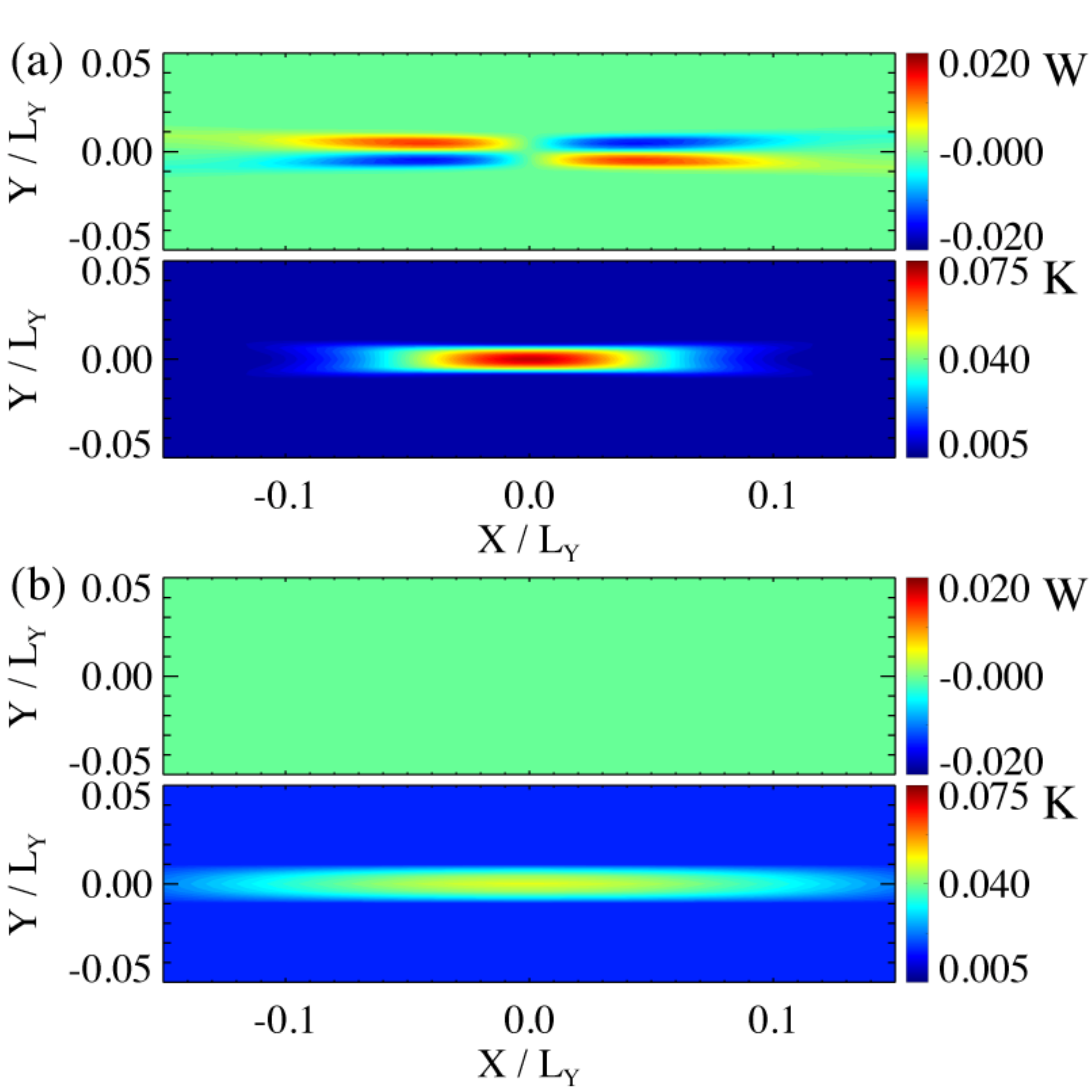}
\caption{\label{fig:turb_ch_effect} Spatial distributions of the turbulent cross helicity (above) and the turbulent MHD energy (below) around the magnetic reconnection  region (a) with the turbulent cross helicity evolution ($W \ne 0$) and (b) with the turbulent cross helicity artificially put equal to zero ($W \equiv 0$).}
\end{figure}

	In order to specify the effects of the cross helicity, we perform a simulation with the turbulent cross helicity artificially put equal zero identically ($W \equiv 0$). Since the turbulent cross helicity is identically zero everywhere, we have no distribution of $W$ at all. Figure~\ref{fig:turb_ch_effect} is for the comparison between (a) the case with cross helicity evolution and (b) the case with the cross helicity identically zero. The spatial distribution of the turbulent MHD energy $K$ with the condition $W \ne 0$ [Fig.~\ref{fig:turb_ch_effect}(a)] is much narrower than the counterpart in the case with $W = 0$ [Fig.~\ref{fig:turb_ch_effect}(b)]. This suggests the cross helicity contributes to the localization of the large magnetic diffusivity region. This tendency itself is natural from the evolution equation of $K$ [Eq.~(\ref{eq:K_eq_num})]. The third term in the r.h.s.\ of Eq.~(\ref{eq:K_eq_num}) is a contribution from the turbulent cross helicity. The sign of $W$ is considered to be the same as that of $\mbox{\boldmath$\Omega$} \cdot {\bf{J}}$. Then $W \mbox{\boldmath$\Omega$} \cdot {\bf{J}} $ is always positive, and the third term of Eq.~(\ref{eq:K_eq_num}) reduces the generation of $K$. However, this reduction vanishes in the immediate vicinity of the symmetry point where the cross helicity vanishes. This contributes to the localization of the turbulent magnetic diffusivity near the X point as we see from the $K$ profile in Fig.~\ref{fig:turb_ch_effect}(a). At the same time, due to the strong inhomogeneity of $W$ near the reconnection point, the fourth term ${\bf{B}} \cdot \nabla W$ reduces the generation of turbulence there. In the close vicinity of the reconnection point, we have no mean magnetic field in the direction of $\nabla W$. As this result, a strong energy production is concentrated in the close vicinity of the reconnection region in the presence of the turbulent cross helicity ($W \ne 0$). 

%------------------------------------------
%	Fig. 7
%------------------------------------------

	One of the prominent features of the present results is the realization of the Petschek type reconnection as a steady state. Localization of $\beta$ due to the turbulent cross helicity contributes to the fast reconnection through the realization of the Petschek reconnection configuration. This validates the theoretical prediction on the cross-helicity effect in magnetic reconnection.\cite{yok2011b}

%-----------------------------------------------------------------------------
%	VII Cross helicity and its effects
%-----------------------------------------------------------------------------
\section{Cross helicity and its effects}
	In the present work, in addition to the enhanced magnetic diffusivity due to the turbulent MHD energy, the transport suppression due to the turbulent cross helicity is considered to play a key role in enhancing the reconnection rate of the mean magnetic field. The cross helicity itself does not enhance the magnetic diffusivity, but works for suppressing the effective transport. It is this property of the cross helicity that works for the localization of the turbulent magnetic diffusivity, leading to enhancement of the magnetic reconnection. In this sense, properties of the cross helicity and its effects are very important. Here we give some notes on the turbulent cross-helicity generation and the cross effect in transport suppression. For further discussions, the reader is referred to Refs.~\onlinecite{yok2011a,yok2011b,yok2013}.

%-----------------------------------------------------------------------------
%	VII.A Turbulent cross-helicity generation
%-----------------------------------------------------------------------------
\subsection{Turbulent cross-helicity generation}
	Asymmetry of the Alfv\'{e}n waves propagating in the opposite directions along the mean magnetic field is one of the main physical sources for the cross-helicity generation. However, if we consider the cross helicity in the context of turbulence, we see that the cross helicity contains some aspects other than the asymmetry of Alfv\'{e}n-wave propagation. This point can be easily understood by looking at the transport equation for the turbulent cross helicity [Eq.~(\ref{eq:K-W_eq})]. Among the several terms which is related to the turbulent cross helicity generation, the first term in Eq.~(\ref{eq:T_W_def}) corresponds to the asymmetry of the Alfv\'{e}n waves. On the other hand, the production terms in Eq.~(\ref{eq:P_W_def}) represent how much cross helicity is cascaded from the mean-field cross helicity. Note that we have exactly the same terms but with the opposite signs in the transport equation for the mean-field cross helicity ${\bf{U}} \cdot {\bf{B}}$. This means that through the mean-field inhomogeneity, the turbulent cross helicity is transferred between the large and small scales. For detailed discussions on the cross-helicity generation mechanisms, see Sec.~5 of Ref.~\onlinecite{yok2013}.

%-----------------------------------------------------------------------------
%	VII.B Cross-helicity effects in transport suppression
%-----------------------------------------------------------------------------
\subsection{Cross-helicity effects in transport suppression}
	The effects of turbulent cross helicity on the transport suppression are twofold. The first one is counterbalancing the turbulent magnetic diffusivity and viscosity with the cross-helicity-related terms. In the mean magnetic induction equation, the turbulent cross helicity represents the transport coefficient $\gamma$ coupled with the mean vorticity (antisymmetric part of the mean velocity shear), $\gamma \mbox{\boldmath$\Omega$}$. In the mean momentum equation, the cross helicity represents the transport coefficient $\nu_{\rm{M}}$ coupled with the mean magnetic-field strain (symmetric part of the mean magnetic shear), $\nu_{\rm{M}} \mbox{\boldmath${\cal{M}}$}$. This suppression effect (in the turbulent EMF) has shown in Fig.~\ref{fig:turb_confinement}. At first glance, the localization due to the turbulent cross helicity seem to be not so great if we see Fig.~\ref{fig:turb_confinement}(a) for $- \beta {\bf{J}}$ and Fig.~\ref{fig:turb_confinement}(c) for $- \beta {\bf{J}} + \gamma \mbox{\boldmath$\Omega$}$. However, there is a notable difference between these two. Due to the spatial distributions of the turbulent cross helicity ($\gamma$) and the mean vorticity $\mbox{\boldmath$\Omega$}$, the transport suppression effect is eminent in the region surrounding the central region. As this consequence, the shapes of Fig.~\ref{fig:turb_confinement}(c) is fairly different in the surrounding region from the counterpart of Fig.~\ref{fig:turb_confinement}(a).  Transport suppression there might have more important meanings if we adopt a more elaborated turbulence model including the density-fluctuation effects. In the surrounding region, the shock--turbulence interaction plays an important role. There the large gradient of the mean density associated with the shocks produces a high level of turbulence. So, the transport suppression in the surrounding region is expected to work well in the localization of transport, which contributes to a fast reconnection. The examination of the transport suppression in the shock--turbulence interaction region is a very interesting subject in the future study.

	The other effect of cross helicity is related to the localization of the turbulent magnetic diffusivity through the localized production of the turbulent MHD energy. As has been already discussed in relation to Fig.~\ref{fig:turb_ch_effect} in \S~VI.C, the turbulent cross helicity has effects on the spatiotemporal evolution of the turbulent MHD energy. These effects are represented by the third and fourth terms on the r.h.s.\ of Eq.~(\ref{eq:K_eq_num}), $- C_\gamma \tau W \mbox{\boldmath$\Omega$} \cdot {\bf{J}}$ and ${\bf{B}} \cdot \nabla W$. Among them, $- C_\gamma \tau W \mbox{\boldmath$\Omega$} \cdot {\bf{J}}$ is the cascade-related effect. The other, ${\bf{B}} \cdot \nabla W$, the cross-helicity inhomogeneity along the mean magnetic field, is related to the Alfv\'{e}n waves.  The gradient of the turbulent cross helicity suggests the inhomogeneity of the Alfv\'{e}n-wave propagations. We consider a domain with Alfv\'{e}nic fluctuations. If the number of outward propagating Alfv\'{e}n-wave packets exceeds the counterpart of the inward propagating ones, the turbulent energy represented by the Alfv\'{e}nic fluctuations will decrease. In this way, the inhomogeneity of the turbulent cross helicity coupled with the mean magnetic field influences the turbulent MHD energy evolution. At the very point of the reconnection, we do not have this effect due to the absence of the mean magnetic field there. But in the region surrounding the reconnection, we have the inhomogeneity of the turbulent cross helicity, then also have the turbulent energy reduction, leading to the localization of the turbulent magnetic diffusivity.

%-----------------------------------------------------------------------------
%	VIII Concluding remarks
%-----------------------------------------------------------------------------
\section{Concluding remarks}
In this work, turbulence effects in the magnetic reconnection environment were explored. Effects of turbulence are incorporated into the mean-field equations through the turbulence correlations such as the Reynolds stress and the turbulent electromotive force. Turbulent transport coefficients appearing in the expressions for the turbulence correlations are expressed in terms of turbulence quantities, and the evolution equations of these turbulence quantities are also considered simultaneously. 

	With the aid of this self-consistent turbulence model, the turbulence effects in the magnetic reconnection were analyzed. In this work, the turbulence is self-generated through the mean-field inhomogeneities and the level and spatiotemporal distribution of turbulence are determined by the dynamics of the mean-field configuration and turbulence itself. Another important aspect in the present work is that the property of magnetic reconnection is determined by the dynamic balance between the transport enhancement due to the turbulent energy and the suppression due to the turbulent cross helicity. The basic role of the turbulent cross helicity is suppressing the enhancement of turbulent transport. Because of this suppression effect, it contributes to confining the region of turbulent magnetic diffusion within a narrow region, leading to the fast reconnection. This is quite different than the current turbulent reconnection model where turbulence leads to a very broad sheet.\cite{laz1999} This difference is attributed to the implementation of the mean-field inhomogeneity in the present work. Without the turbulence (energy, cross helicity, etc.) generation due to the mean-field inhomogeneity, it would be very difficult to realize a localization of the reconnection region. The present numerical simulation supports this scenario of transport enhancement and suppression at the mean-field level.
	
	To summarize, the present results suggest:
\begin{itemize}
\item
Turbulence (turbulent energy and turbulent cross helicity) is self-generated and sustained by the inhomogeneity of the mean fields.

\item
Just reflecting the symmetry of the mean-field configurations, the turbulent cross helicity is spatially distributed in a quadrupole-like manner around the reconnection point.

\item
Turbulent magnetic diffusivity is balanced by the cross helicity effect which is acting for the field generation or transport suppression.

\item
Since the turbulent cross helicity vanishes at the symmetric point, the effect of the turbulent magnetic diffusivity is dominant only in the vicinity of such symmetric point, i.e., reconnection point.

\item
The turbulent cross helicity contributes to the confinement of the turbulent magnetic diffusion, leading to the fast reconnection.
\item
If effects of turbulence are incorporated properly (in a self-consistent manner with considering balance between turbulent transport enhancement and suppression), even in the framework of the one-fluid MHD, a fast reconnection can be realized as a steady state at the mean-field level.

\end{itemize}

	The numerical simulation adopted in this work is based on a turbulence model. Further numerical tests for this notion of dynamic balance in magnetic reconnection with direct numerical simulations (DNSs) would be a very interesting subject to investigate. In such tests with DNSs, how to generate and sustain turbulence may be a key issue. In the present work, turbulence is self-generated and -sustained by the inhomogeneities of the mean fields. This point is in strong contrast to the previous numerical simulation work with fully-developed turbulence.\cite{ser2009,kow2009,kow2012,lou2009} At the same time, further validation of the theoretical expressions for the turbulent electromotive force ${\bf{E}}_{\rm{M}}$ [Eq.~(\ref{eq:emf_exp})] and the Reynolds stress $\mbox{\boldmath${\cal{R}}$}$ [Eq.~(\ref{eq:rey_stress_exp})] by DNSs is required. 

	In the numerical simulation in this work, we perform a two-dimensional calculation for the sake of simplicity. This is the main reason why we dropped the helicity-related effect ($\alpha$ and $\mbox{\boldmath$\Gamma$}$) in the model equation. However, this does not deny the importance of the kinetic, current, and magnetic helicities in general situations. Actually in the three-dimensional reconnection situations, we have non-zero helicities in the mean-field configuration, then naturally have non-zero helicities also in turbulence, which may participate in the dynamic balance of turbulent transport. In such  situations, we have to take helicities other than the cross helicity into account. This gives very interesting subjects to survey.

	The spatial distributions of the pseudoscalar turbulence quantities, in the present case the quadrupole distribution of the turbulent cross helicity, reflect just the symmetric and antisymmetric properties of the mean-field configurations in the magnetic reconnection environment. In this sense, such spatial distributions must be very ubiquitous around the reconnection point. It is still very difficult to simultaneously measure three components of both magnetic field and velocity in remote observations. However, in in-situ observations with probes, such simultaneous measurement can be performed much easier. It is hard to expect that a satellite observation is operated at the very point of the magnetic reconnection in the magnetosphere. But, if a spacecraft passes nearby the point of reconnection, the turbulent cross helicity (correlation between the velocity and magnetic-field fluctuations) should be observed to change its sign due to the quadrupole spatial distribution of the turbulent cross helicity. The experimental and/or satellite observational measurements of the cross helicity around the reconnection point would be very interesting subjects. Since the sign of cross helicity should change across the neutral or symmetric plane, even the detection of the sign itself would give very interesting information.
	
	Another interesting point is related to external sources of the cross correlation between the velocity and magnetic-field fluctuations. As we have seen, the presence of turbulent cross helicity leads to the change in the transport property of turbulence. In the present work, the turbulent cross helicity, as well as the turbulent energy, is generated and sustained by the nonlinear dynamics of the mean-field configurations with turbulence itself. However, in some situations, we may have other ``external'' sources that provide the cross helicity in turbulence; asymmetry of the Alfv\'{e}n-wave propagation, neutral beam injection (NBI) with shear in fusion devices, polarization current etc. There may be some beyond the description of usual MHD. The cross helicity in turbulence is expected to contribute to the dynamic balance in turbulent transport, no matter what may be the source of the cross helicity. In this sense, the cases with external cross-helicity sources would provide interesting subjects to explore.
%\vspace{10pt}

%-----------------------------------------------------------------------------
%	Acknowledgments
%-----------------------------------------------------------------------------
\begin{acknowledgments}
The authors would like to thank the anonymous referees for their useful comments which improved the presentation of this paper much. Their thanks are also due to invaluable comments by Gene Parker, Annick Pouquet, Amitava Bhattacharjee, Jim Drake, J\"{o}rg B\"{u}chner, and Homa Karimabadi. Part of this work was performed during the period when one of the authors (N.Y.) stayed at the Nordic Institute for Theoretical Physics (NORDITA) in October, at the Princeton Plasma Physics Laboratory (PPPL) in November, and at the Department of Physics, University of Wisconsin-Madison in December 2012 as a visiting researcher. This work is supported by the Japan Society for the Promotion of Science (JSPS) Core-to-Core Program (No.~22001) Institutional Program for Young Researcher Overseas Visits and also by the JSPS Grants-in-Aid for Scientific Research (No.~24540228).
\end{acknowledgments}

%------------------------------------------------------------------------------
%	References
%------------------------------------------------------------------------------


\begin{thebibliography}{99}

\bibitem{swe1958}
P. A. Sweet,
``The neutral point theory of solar flares,''
in IAU Symp.\ {\bf{6}}, {\it Electromagnetic Phenomena in Cosmical Plasma}, 
ed.\ B. Lehnert (Cambridge University Press, New York, 1958), 123-134.

\bibitem{par1957}
E. N. Parker,
``Sweet's mechanism for merging magnetic fields in conducting fluids,''
J. Geophys.\ Res.\ {\bf{62}}, 509-520 (1957).

\bibitem{pet1964}
H. E. Petschek,
``Magnetic field annihilation,''
in AAS-NASA Symposium (NASA SP-50), 
ed.\ W. H. Hess (NASA, Greenbelt, MD), 425-439 (1964).

\bibitem{vas1975}
V. M. Vasyliunas,
``Theoretical models of magnetic field line merging. I,''
Rev.\ Geophys.\ Space Phys.\ {\bf{13}}, 303-336 (1975).

\bibitem{ter1983}
T. Terasawa,
``Hall current effect on tearing mode instability,''
Geophys.\ Res.\ Lett.\ {\bf{10}}, 475-478 (1983). 

\bibitem{bir2001}
J. Birn,
``Geospace Environmental Modeling (GEM) magnetic reconnection challenge,''
J. Geophys.\ Res.\ {\bf{106}}, 3737-3750 (2001).

\bibitem{hes1998}
M. Hesse and D. Winske,
``Electron dissipation in collisionless magnetic reconnection,''
J. Geophys.\ Res.\ {\bf{103}}, 26479-26486 (1998).

\bibitem{sin2011}
K. A. P. Singh, K. Shibata, N. Nishizuka, and H. Isobe,
``Chromospheric anemone jets and magnetic reconnection in partially ionized
solar atmosphere,''
Phys.\ Plasmas {\bf{18}}, 111210-1-8 (2011).

\bibitem{buc1999}
J. B\"{u}chner,
``Three-dimensional magnetic reconnection in astrophysical plasmas -- Kinetic approach,'' 
Astrophys.\ Space Sci.\ {\bf{264}}, 25-42 (1999).

\bibitem{hig2013b}
K. Higashimori and M. Hoshino,
``Self-generation of turbulence in collisionless magnetic reconnection:
ion plasma beta dependence,''
Phys.\ Plasmas (to be submitted).

\bibitem{hig2013a}
K. Higashimori, N. Yokoi, and M. Hoshino,
``Explosive turbulent magnetic reconnection,''
Phys.\ Rev.\ Lett.\ {\bf{110}}, 255001-1-5 (2013).

\bibitem{hos1994}
M. Hoshino, A. Nishida, T. Yamamoto, and S. Kokubun,
``Turbulent magnetic field in the distant magnetotail: Bottom-up process of plasmoid formation?"
Geophys.\ Res.\ Lett.\ {\bf{21}}, 2935-2938 (1994).

\bibitem{taj1997}
T. Tajima and K. Shibata,
{\it Plasma Astrophysics},
(Addison--Wesley, Reading, 1997).

\bibitem{kar2013}
H. Karimabadi and A. Lazarian,
``Magnetic reconnection in the presence of externally driven and self-generated turbulence,''
Phys.\ Plasmas {\bf{20}}, 112102-1-16, (2013).

\bibitem{mat1985}
W. H. Matthaeus and S. L. Lamkin,
``Rapid magnetic reconnection caused by finite amplitude fluctuations,''
Phys.\ Fluids {\bf{28}}, 303-307 (1985).

\bibitem{mat1986}
W. H. Matthaeus and S. L. Lamkin,
``Turbulent magnetic reconnection,''
Phys.\ Fluids {\bf{29}}, 2513-2534 (1986).

\bibitem{gol1995}
P. Goldreich and S. Sridhar,
``Toward a theory of interstellar turbulence: II. Strong alfv\'{e}nic turbulence,''
Astrophys.\ J. {\bf{438}}, 763-775 (1995).

\bibitem{laz1999}
A. Lazarian and E. T. Vishniac,
``Reconnection in a weakly stochastic field,''
Astrophys.\ J. {\bf{517}}, 700-718 (1999).

\bibitem{eyi2011}
G. L. Eyink, A. Lazarian, and E. T. Vishniac,
``Fast magnetic reconnection and spontaneous stochasticity,''
Astrophys.\ J. {\bf{743}}, 51-1-28 (2011).

\bibitem{guo2012}
Z. B. Guo, P. H. Diamond, and X. G. Wang,
``Magnetic reconnection, helicity dynamics, and hyper-diffusion,''
Astrophys.\ J. {\bf{757}}, 173-1-15 (2012).

\bibitem{ser2009}
S. Servidio, W. H. Matthaeus, M. A. Shay, P. A. Cassak, and P. Dmitruk,
``Magnetic reconnection in two-dimensional magnetohydrodynamic turbulence,''
Phys.\ Rev.\ Lett.\ {\bf{102}}, 115003-1-4 (2009).

\bibitem{kow2009}
G. Kowal, A. Lazarian, E. T. Vishniac, and K. Otmianowska-Mazur,
``Numerical tests of fast reconnection in weakly stochastic magnetic fields,''
Astrophys.\ J. {\bf{700}}, 63-85 (2009).

\bibitem{kow2012}
G. Kowal, A. Lazarian, E. T. Vishniac, and K. Otmianowska-Mazur,
``Reconnection studies under different types of turbulence driving,''
Nonlin.\ Processes Geophys.\ {\bf{19}}, 297-314 (2012).

\bibitem{lou2009}
N. F. Loureiro, D. A. Uzdensky, A. A. Shekochihin, S. C. Cowley, and T. A. Yousef,
``Turbulent magnetic reconnection in two dimensions,''
Mon.\ Not.\ Roy.\ Astron.\ Soc.\ {\bf{399}}, L146-1-5 (2009).

\bibitem{dau2009b}
W. Daughton, V. Roytershteyn, B. J. Albright, H. Karimabadi, L. Yin, and K. J. Bowers,
``Transition from collisional to kinetic regimes in large-scale reconnection layers,''
Phys.\ Rev.\ Lett.\ {\bf{103}}, 065004-1-4 (2009).

\bibitem{uzd2010}
D. A. Uzdensky, N. F. Loureiro, and A. Schekochihin,
``Fast magnetic reconnection in the plasmoid-dominated regime,''
Phys.\ Rev.\ Lett.\ {\bf{105}}, 235002-1-4 (2010).

\bibitem{nak2013}
T. Nakabo, K. Kusano, T. Miyoshi, and G. Vekstein, 
``Simulation study of magnetic reconnection in high magnetic Reynolds number plasmas,'' 
American Geophysical Union, Fall Meeting 2013, abstract SM13B-2139 (2013).

\bibitem{bha2009}
A. Bhattacharjee, Yi-Min Huang, H. Yang, and B. Rogers,
``Fast reconnection in high-Lundquist-number plasmas due to the plasmoid instability,''
Phys.\ Plasmas {\bf{16}}, 112102-1-5 (2009).

\bibitem{dau2009a}
W. Daughton, V. Roytershteyn, B. J. Albright, H. Karimabadi, L. Yin, and K. J. Bowers,
``Influence of Coulomb collisions on the structure of reconnection layers,''
Phys.\ Plasmas {\bf{16}}, 072117-1-16 (2009).

\bibitem{ji2011}
H. Ji and W. Daughton,
``Phase diagram for magnetic reconnection in heliophysical, astrophysical, and laboratory plasmas,''
Phys.\ Plasmas {\bf{18}}, 111207-1-10 (2011).

\bibitem{yok2011b}
N. Yokoi and M. Hoshino, 
``Flow--turbulence interaction in magnetic reconnection,'' 
Phys.\ Plasmas {\bf{18}}, 111208-1-14 (2011).

\bibitem{yos1990}
A. Yoshizawa,
``Self-consistent turbulent dynamo modeling of reversed field pinches and planetary magnetic fields,''
Phys.\ Fluids B {\bf{2}}, 1589-1600 (1990).

\bibitem{yok2013}
N. Yokoi, 
``Cross helicity and related dynamos,'' 
Geophys.\ Astrophys.\ Fluid Dynamics {\bf{107}}, 114-184 (2013).

\bibitem{yok2011c}
N. Yokoi and G. Balarac, 
``Cross-helicity effects and turbulent transport in magnetohydrodynamic flow,'' 
J. Phys.\ Conf.\ Ser.\ {\bf{318}}, 072039-1-10 (2011).

\bibitem{zho1990}
Y. Zhou and W. H. Matthaeus,
``Transport and turbulence modeling of solar wind fluctuations,''
J. Geophys.\ Res.\ A {\bf{95}}, 10291-10311 (1990).

\bibitem{tum1993}
C.-Y. Tu and E. Marsch,
``A model for solar wind fluctuations with two components:
Alfv\'{e}n waves and convective structures,''
J. Geophys.\ Res.\ A, {\bf{98}}, 1257-1276 (1993).

\bibitem{yok2007}
N. Yokoi and F. Hamba,
``An application of the turbulent magnetohydrodynamic
residual-energy equation model to the solar wind,''
Phys.\ Plasmas, {\bf{14}}, 112904-1-16 (2007).

\bibitem{yok2011a}
N. Yokoi, 
``Modeling turbulent cross helicity evolution: Production, dissipation, and transport rates,'' 
J. Turbulence {\bf{12}}, N27-1-33 (2011).

\bibitem{bir2007}
J. Birn and E. R. Priest,
{\it Reconnection of Magnetic Fields},
(Cambridge University Press, Cambridge, 2007).

\bibitem{yos2013}
A. Yoshizawa, H. Kobayashi, N. Sugimoto, N. Yokoi, and Y. Shimomura,
``A Reynolds-averaged turbulence modeling approach to the maintenance of the Venus superrotation,''
Geophys.\ Astrophys.\ Fluid Dynamics {\bf{107}}, 614-639 (2013).

\bibitem{mar2001}
J. Maron and P. Goldreich,
``Simulations of incompressible magnetohydrodynamic turbulence,''
Astrophys.\ J. {\bf{554}}, 1175-1196 (2001).

\bibitem{cho2002}
J. Cho, A. Lazarian, and E. T. Vishniac,
``Simulations of magnetohydrodynamic turbulence in a strongly magnetized medium,''
Astrophys.\ J. {\bf{564}}, 291-301 (2002).

\bibitem{ber2008}
A. Beresnyak and A. Lazarian,
``Strong imbalanced turbulence,''
Astrophys.\ J. {\bf{682}}, 1070-1075 (2008).

\bibitem{men1996}
M. Meneguzzi, H. Politano, A. Pouquet, and M. Zolver,
``A sparse-mode spectral method for the simulation of turbulent flows,''
J. Comp.\ Phys.\ {\bf{123}}, 32-44 (1996).

\bibitem{bol2006}
S. Boldyrev,
``Spectrum of magnetohydrodynamic turbulence,''
Phys.\ Rev.\ Lett.\ {\bf{96}}, 115002-1-4 (2006).

\bibitem{mas2008}
J. Mason, F. Cattaneo, and S. Boldyrev,
``Numerical measurements of the spectrum in magnetohydrodynamic turbulence,''
Phys.\ Rev.\ E {\bf{77}}, 036403 (2008).

\bibitem{yos1996}
A. Yoshizawa, 
``Turbulent magnetohydrodynamic dynamo: Modeling of the turbulent residual-helicity equation,'' 
J. Phys.\ Soc.\ Jpn.\ {\bf{65}}, 124-132 (1996).

\bibitem{yok2008}
N. Yokoi, R. Rubinstein, A. Yoshizawa, and F. Hamba,
``A turbulence model for magnetohydrodynamic plasmas,''
J. Turbulence {\bf{9}}, N37-1-25 (2008).

\bibitem{lau1972}
B. E. Launder and D. B. Spalding, 
{\it Lectures in Mathematical Models of Turbulence},
(Academic Press, London, 1972).

\bibitem{yos1987}
A. Yoshizawa,
``Statistical modeling of a transport equation for the kinetic energy dissipation rate,''
Phys.\ Fluids {\bf{30}}, 628-631 (1987).

\bibitem{ric1926}
L. F. Richardson,
``Atmospheric diffusion shown on a distance-neighbour graph,''
Proc.\ Roy.\ Soc.\ Lon.\ Series A {\bf{110}}, 709-737 (1926).

\bibitem{kra1959}
R. H. Kraichnan,
``The structure of isotropic turbulence at very high Reynolds numbers,''
J. Fluid Mech.\ {\bf{5}}, 497-543 (1959).


\end{thebibliography}
\end{document}